\documentclass[default,iicol]{sn-jnl}

\jyear{2021}%

\theoremstyle{thmstyleone}
\theoremstyle{thmstyletwo}%

\theoremstyle{thmstylethree}%

\usepackage{tabularx}
\usepackage{diagbox}
\usepackage{rotating}
\usepackage{colortbl}
\usepackage{amssymb}
\usepackage{pifont}
\newcommand{\cmark}{\ding{51}}%
\newcommand{\xmark}{\ding{55}}%

\begin{document}

\bibliographystyle{bst/sn-aps}

\title[]{Edge Computing: A Systematic Mapping Study}


\author[1]{\fnm{Jalal} 
\sur{Sakhdari}}\email{jalalsakhdari@mail.um.ac.ir}

\author[1]{\fnm{Behrooz} 
\sur{Zolfaghari}}\email{be.zolfaghari@mail.um.ac.ir}

\author[1]{\fnm{Shaghayegh} 
\sur{Izadpanah}}\email{sh.izadpanah@mail.um.ac.ir}

\author[1]{\fnm{Samaneh} 
\sur{Mahdizadeh Zargar}}\email{smahdizadeh@um.ac.ir}

\author[1]{\fnm{Mahla} 
\sur{Rahati Quchani}}\email{mahla.rahati@mail.um.ac.ir}

\author[1]{\fnm{Mahsa} 
\sur{Shadi}}\email{mahsa.shadi@mail.um.ac.ir}

\author*[1]{\fnm{Saeid} 
\sur{Abrishami}}\email{s-abrishami@um.ac.ir}

\author[1]{\fnm{Abbas} 
\sur{Rasoolzadegan}}\email{rasoolzadegan@um.ac.ir}

\affil[1]{\orgdiv{Department of Computer Engineering}, \orgname{Ferdowsi University}, \orgaddress{\street{Azadi Street}, \city{Mashhad}, \country{Iran}}}


\abstract{Edge computing is a novel computing paradigm that extends cloud resources at the edge of the network to tackle the problem of communication latency in latency-sensitive applications. For the last decades, there have been many efforts dedicated to this field. The dramatic growth in the publications, and the great attention of the research community, have made it necessary to conduct a Systematic Mapping Study (SMS) to enable researchers get a better view of the field. A three-tier search method is considered in this work. In this method, we defined some quality criteria to extract appropriate search spaces and studies. Using this methodology, we select 112 search spaces out of 805 ones, and by searching in these search spaces we select 1440 studies out of 8725. In our SMS, 8 research questions have been designed and answered to identify the main topics, architectures, techniques, etc. in the field of edge computing.}




\maketitle

\section{Introduction}\label{sec1}

By increasing the number of internet-based applications and their heavy computation or a large amount of data, remote computational resources offered by data centers of the Cloud Computing (CC) paradigm is a powerful solution to data storage and processing requirements. In fact, the computation or data of these applications should be transferred to data centers of clouds, which are geographically established far from users running applications. Due to this distance, and consequently, its huge communication latency, this computation model is not appropriate for applications that need mobility support, location awareness, and low latency.

To solve the above problem, edge computing has appeared as an alternative computing model. Its main idea is to bring computing and storage resources closer to the end-users and at the edge of the network, which leads to a considerable reduction in communication latency. The field of edge computing has been significantly investigated for around two decades. The concept was firstly introduced by Akamai by creating cache servers on the edge of the network \cite{r1}. This idea has also been used in P2P systems to distribute tasks and workloads between peers. However, the concept of edge computing has evolved, and various architectures have been introduced to it.

There are several definitions in the literature for edge computing \cite{r2,r3,r4}. In this SMS, we have reviewed the researches in which processing or storage processes are sent to devices close to the user, in the category of edge computing. Although edge computing is a promising way to meet the quality needs of new applications, there are still many operational challenges. On the other hand, considerable growth has been observed in researches in the field of edge computing due to the increasing use of IoT applications and the significant increase in mobile and wearable devices. Besides, due to the large volume of papers published in this field, especially in recent years, conducting a survey on the literature can help researchers to identify the core scopes of edge computing. In other words, we need a guide to enable researchers to do more effective searches on each scope of the edge computing. Due to the vastness and number of researches in the field of edge computing, we need an advanced research method. This method should be able to cover research works in the field of edge computing, and unbiased, rigorous and auditable (traceable). To the best of our knowledge, although there are many research studies that survey existing works, none of them uses a standard and systematic method for searching and reviewing papers.

To conduct an advance and comprehensive literature review, evidence-based software engineering recommends two well-known research techniques as Systematic Literature Review (SLR) and Systematic Mapping Study (SMS) \cite{r5,r6}. Despite the same search and data extraction methodology of these two systematic reviews, they have some differences in their goal, which distinguish their usage \cite{r7}. In SMS, researchers try to gain general information in a specific research field. More precisely, they aim to classify topics and identify trends existing in that research field, without paying special attention to the details of each paper. On the other hand, SLR investigates the primary subset of research studies and tries to extract more specific data from them in order to analyse a topic (sub-topic) in-depth and mention the advantages and shortcuts of different proposals. This difference arises from different Research Questions (RQs) in each of these two techniques. The research questions for an SMS are quite high level i.e., which sub-topics have been addressed and provide an overview of the literature in specific topic areas \cite{r5}. Hence, SMS should be applied when identifying the scopes, and categorizing a large number of studies is considered, while SLR should be utilized when specific data of a limited number of studies are needed. It should be noted that SMS can be performed as a pre-SLR review.    

To conduct an SMS, a systematic method is needed so that we can extract the search spaces and research studies published in the field of edge computing and then select some of them with appropriate quality. In this paper, the known methodology presented in \cite{r5,r6} has been used. The central part of this methodology is to provide an appropriate search method to extract a large volume of related studies in the field of edge computing. For this purpose, related studies have been searched at three levels (Manual search, Backward snowballing, and Database search). We have also defined some quality criteria to select search spaces and studies with acceptable quality and maximum relevance to the field of edge computing, among all the extracted items for further analysis. Finally, in a separate phase, we evaluated the extraction process of related studies in terms of accuracy and collecting the comprehensive and complete information in a supplementary file (Edge\_SupportFile file available at  \url{https://github.com/jalalsakhdari/SMS}). In this SMS, we have used the experiences and feedbacks from our past systematic reviews \cite{r7,r8,r9}and tried to provide a better review.

In this work, we apply a SMS on the field of edge computing that explore the edge computing topic. We have classified the topics existing in the edge computing field, and also categorized researched studies within these topics. Our ultimate goal is to effectively alleviate specifying research trends in the terms of topics and sub-topics (as done in an SMS \cite{r5})for researchers interested in this field. 

To reach this goal, 8 research questions have been designed, and we tried to answer and analyze these questions during the systematic review process. These questions are designed to achieve goals such as identifying critical topics in the field of edge computing, identifying architectures used, identifying commonly used techniques, applications of edge computing in other areas, identifying active researchers, and active search spaces in the field of edge computing. This information will be a useful guide for researchers and developers interested in the field of edge computing.

The rest of this paper is structured as follows: In section 2, the steps of an SMS process are detailed, followed by research methodology applied in this paper, and the threats to validity of this review. In section 3, results achieving from our SMS are presented and discussed. In section 4, we compare our SMS with other secondary studies. In section 5, we present some implications based on the results of our SMS for researchers, practitioners, and teachers interested in the field under study.  Finally, in section 6, we conclude our review. 

Along with this study, for more clarity, we have provided a comprehensive supplementary document to provide details of our findings (which is accessible from https://github.com/jalalsakhdari/SMS). Throughout the paper, we will reference specific sections of this documents that will be useful to readers interested in the details of the method and the data extracted. These documents have the prefix SupFile, plus a postfix which specifics a table. For example, if we reference [SupFile]\_(E1,T1), it refers to Table T1 of the Excel file E1. Also, for more clarity, all abbreviations used in this paper are summarized in Table\ref{tbl1}.

\begin{table*}[h]
\begin{center}
\begin{minipage}{\textwidth}
		\caption{Summary of abbreviations.}\label{tbl1}
		
	\begin{tabular*}{\textwidth}{@{\extracolsep{\fill}}llll@{\extracolsep{\fill}}}
			\toprule
			\textbf{Abber.} & \textbf{Term} & \textbf{Abbre.} & \textbf{Term}  \\
			\midrule
	
			EC &  Edge Computing         & D2D  & Device-to-Device \\  
			FC & Fog Computing           & M2M  & Machine-to-Machine \\
			CO & Computation Offloading  & BS   & Base Station \\
			Ar.& Architecture            & SDN  & Software-Defined Network \\
			DM & Data Management         & NFV  & Network Function Virtualization \\
			RM & Resource Management     & TDMA & Time Division Multiple Access \\
			NM & Network Management      & FDMA & Frequency Division Multiple Access \\
			Ec.&  Economic               & CDMA & Code Division Multiple Access \\
			S\&P&  Security \& Privacy     & P2P & Peer to Peer \\
			UE &  User Equipment         & RAN & Radio Access Network \\
			MD &  Mobile Device          & BBU & Base Band Unit \\
			VM &  Virtual Machine        & RRH & Radio Remote Head \\
			RQ &  Research Question      & ANET& Vehicular Ad-hoc Network \\
			MEC & Mobile Edge Computing  & QoS & Quality of Service \\
			MCC & Mobile Cloud Computing & QoE & Quality of Experience \\
			IoT	& Internet of Things     &  -   &   -      \\ 
			\botrule
		\end{tabular*}
		\end{minipage}
		\end{center}
	\end{table*}

\section{Research methodology}\label{sec2}
In this paper, we use the methodology, which is based on the updated SMS guidelines described \cite{r6}. Generally, the SMS process of this paper is comprised of three main phases. (i) The planning phase, (ii) Evaluating phase, and (iii) Conducting phase. In the following, due to space constraint, each of these steps are listed briefly in here and in Appendix A (Appendix A\_v1 file available at \url{https://github.com/jalalsakhdari/SMS}) has been completely explained.

\begin{enumerate}
		\itemsep=0pt
		\item Specifying the Scope and RQs (each RQ, along with its rationality, is summarized in Table\ref{tbl2})
		\item Planning of the search strategy including specify the search strategy, search space, and search strings
		\item Specifying the search Space
		\item Specifying the search string 
		\item	Planning the study selection process
		\item Specifying the search and study selection evaluation strategy
		\item Planning of the data extraction and classification process
	\end{enumerate}
	
\begin{figure*}[t]
		\centering
		\includegraphics[width=\textwidth]{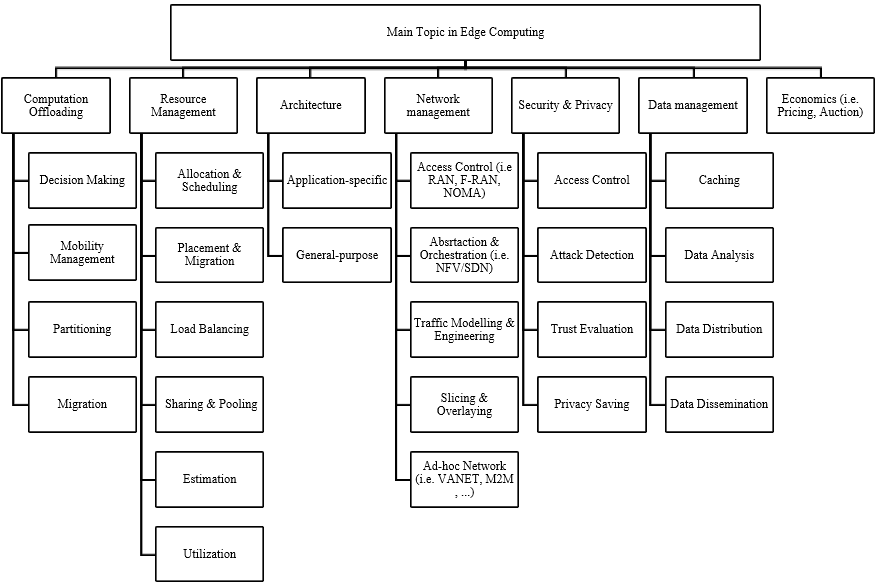}
		\caption{Edge Computing Research Tree.}
		\label{fig:1}
	\end{figure*}

After completing above step and identifying the included studies, to respond to the RQs of the SMS, these data should accurately be investigated and analyzed. The most important objective of this SMS is identifying the primary topics and sub-topics in the field of edge computing. Figure\ref{fig:1} shows the research tree obtained during this SMS. How to extract the research tree is fully described in Appendix A. 

During the search process (including planning phase or conducting phase), various factors may affect the validity of SMS. Therefore, one of the crucial rubrics is the discussion of threats to validity. The primary purpose of the validation process is to provide evidence to respond to all the threats that the systematic review process may face. Some of these salient evidences are discussed and presented in Appendix A.

\begin{table*}[h]
\begin{center}
\begin{minipage}{\textwidth}
		\caption{Defined Research Questons(RQs).}\label{tbl2}
	\begin{tabular*}{\textwidth}{@{\extracolsep{\fill}} p{0.2cm} p{4cm} p{10cm} @{\extracolsep{\fill}}}
		\toprule
			RQ & Research Questions & Rationale \\
			\midrule
			1 & Which researchers and research venues are more active in this field and how are the active researchers distributed geographically? & The demographics of edge computing research provide a useful starting point for interested researchers by identifying active scholars, venues, and countries. \\
			2 & How active is the field of edge computing and how is the distribution of selected studies by type over publication year and Geographical areas?  & To identify the current volume of research and general trends in order to better depict the attractiveness of the field. Comparing the volume of research over publication year and geographical areas can shed some light on the maturity of edge computing \\
			3 & What are the core research topics in the field of edge computing?  & To identify and classify the current research topics in the field of edge computing, the evaluation and distribution of each topic, and the potential trends \\
			4 & What is the distribution of applications in each research topics?	& To identify the percentage of applicability and the importance of each research topics based on the number of studies published in each application relative to the total number of studies. \\
			5 & What is the distribution of different architectures in each topic? &	Architecture can be Fog Computing, Cloudlet, MEC and so on. \\
			6 & Which techniques are more used in the field? &	To identify main techniques used in each topic of edge computing field and analyzing the relation between techniques with other aspects. Techniques can be game theory, heuristic and so on \\
			7 & Which forms of empirical evaluation have been used?	& Empirical evaluation means whether the environment is real or simulation or testbed.  \\
			8 & Mostly, Which Qualitative Requirements (QoS) have been considered to move towards edge computing? &	Answering this research question will make it possible to understand when a researcher must use edge computing. QoS can be time, cost, energy, and so on. \\
			\botrule
		\end{tabular*}
		\end{minipage}
		\end{center}
	\end{table*}
	
	\section{Results of the study}
	In the previous sections, the search process and its details were described. In this section, based on the research questions (RQs) posed in Section 2, the results of reviewing the field of edge computing are discussed. The RQs are answered in the order in which they appear in Table\ref{tbl2}. Furthermore, questions are answered and reviewed based on the levels of the research tree.
	
	\subsection{How active is the field of edge computing, and how is the distribution of selected studies by type over publication year (journal) and Geographical areas?}
	
\begin{figure}[t]
		\centering
		\includegraphics[scale=0.51]{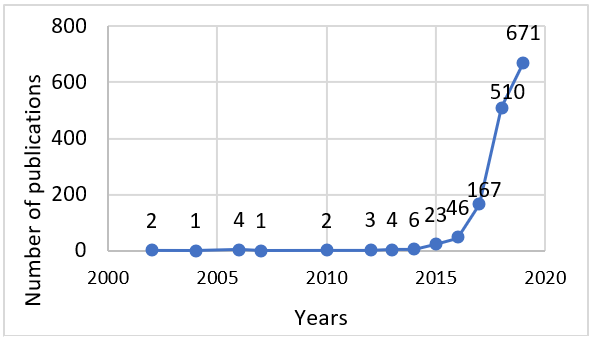}
		\caption{Number of Publications per Year}
		\label{fig:2}
	\end{figure}

	In this section, the number of publications in different years is analyzed statistically, and the percentage of the number of studies on different topics is informed. Figure\ref{fig:2} shows the number of papers published in the field of edge computing over the years. In this figure, the horizontal axis represents the years, and the vertical axis represents the number of published papers. The information in the [SupFile]\_(E2,T1) and [SupFile]\_(E1,T5) files are used to draw the diagrams in this section.
	
	As shown in Figure\ref{fig:2}, until 2014, few papers have been published in this field. In previous years, similar researches have been done to performing computations close to the user such as cloudlets \cite{r10}, cyber foraging \cite{r11}, and nomadic computing \cite{r12}, but the starting point for the emergence of formal, and standard edge computing can be attributed to 2013-2015. In 2013, IBM, and Nokia unveiled the Joint Radio Applications Cloud Server (RACS) project, a platform for edge computing over 4G / LTE networks. After that, efforts began to standardizing edge computing under the auspices of the European Telecommunications Standards Institute (ETSI) \cite{r13}. The first Mobile Edge Computing Congress was held in London in 2015, followed by the first IEEE / ACM Symposium on Edge Computing a year later. After these events, edge computing became known as a new research path with many research opportunities among researchers, academic, and industrial projects.

Another noticeable point in Figure\ref{fig:2} is the sharp upward trend of publications in recent years. The following facts can be considered as some of the reasons to justify the increasing popularity of this research field, and a significant increase in the number of papers in recent years:

\textit{IoT maturity, and the emergence of new applications:} With the widespread use of smartphones, and the rise of IoT-based technologies, a variety of applications such as healthcare, smart community, social networks, and VANET have become more popular. As a result, a large number of researches have been done to adapt these applications to the edge computing paradigm and improving the quality of these applications through the use of edge, and fog.

\textit{The existence of several operational challenges:} Edge computing is an emerging field of research. Therefore, there are still many challenges in resource management, networking, and QoS assurance in this area. This issue has caused the attention of many researchers in this field.

\textit{Advantages of edge over remote cloud:} With the connection of billions of devices to the internet, the remote cloud will not be operational for new applications. Consequently, it will impose heavy traffic on the backbone network, and cause communication delays \cite{r13}. Thus, a new flood of researches has been done to migrate and adapt remote cloud services and applications to distributed edge resources.

\textit{Close relationship to other research area and technologies:} The field of edge computing is closely related to various areas of research including cloud computing, networking, data analysis and processing, security, artificial intelligence, medicine, etc. This has led many researchers from these different fields tend to publish papers in edge computing.

\begin{figure}[t]
	\centering
	\includegraphics[scale = 0.51]{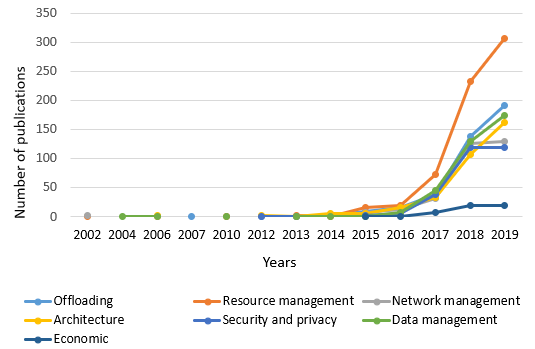}
	\caption{Number of Publications per Topics}
	\label{fig:3}
\end{figure}

Figure\ref{fig:3} shows the number of published papers in the field of edge computing over the years and for each topic, separately. In this figure, the horizontal axis represents the years, and the vertical axis represents the number of published papers. As this figure illustrates, the upward trend in studies publishing is evident for all topics. Among these, the resource management, and computation offloading topics have had the highest acceleration compared to other topics. Since the resources in edge computing are limited and heterogeneous with variable load, resource management is one of the critical issues in this field \cite{r14}. As a result, the number of resource management researches has grown more sharply than other topics. The field of computation offloading is also one of the active research fields with complicated and diverse challenges such as mobility management, partitioning, and decision making, so it has attracted much attention. After resource management, and computation offloading topics, the data management topic has had the highest growth of the number of publications. The reason for this growth can also be attributed to the typical applications in this field. Most edge computing applications such as healthcare, crowdsourcing, streaming and social network are working with user data. Therefore, there is a strong need to provide methods for storing, processing and analyzing data in a distributed edge environment. For this reason, data management has also become one of the fastest-growing topics in this field.

\subsection{Which researchers and research venues are more active in this field, and how are the active researchers distributed geographically?}

In this section, active journal researchers, and journals of edge computing are introduced. The information of this section can be a good starting point for new researchers. To identify active researchers, the number of studies published by each author is counted. The first ten authors with the highest number of researches are selected as active researches, and the number of their publications in each topic are counted. To do this, the information of the authors’ column in the [SupFile]\_(E1,T1) is used. Information on the number of studies published by other authors per topic is also available in the [SupFile]\_(E4,T4). Table\ref{tbl3} depicted a small part of this file. This table shows the number of published papers per topic and the total number of published papers for each author. Since some papers are categorized into more than one topic, the total number of papers published on different topics may be higher than the total number of papers published by a specific author.

\begin{figure}[b]
	\centering
	\includegraphics[scale=0.44]{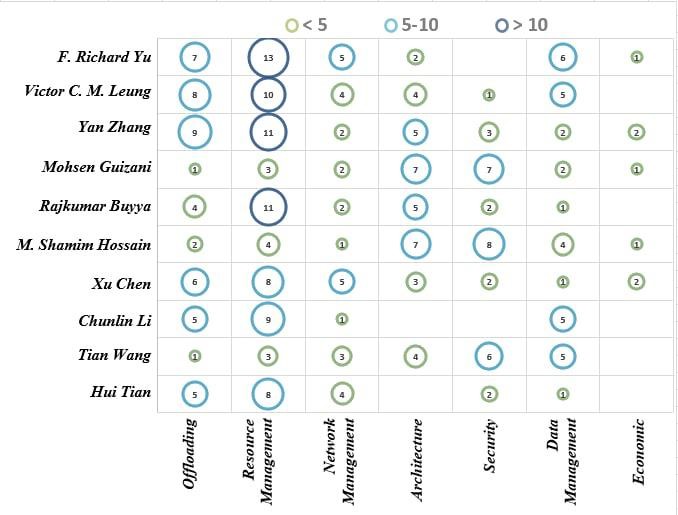}
	\caption{Active Researchers in the Field of Edge Computing}
	\label{fig:4}
\end{figure}

In Figure\ref{fig:4} information about active researchers are shown in the form of a bubble diagram. In this diagram, the horizontal axis represents the topics and the vertical axis represents the top ten authors with the highest number of journal publications and the volume of bubbles represents the number of papers published in each topic. According to the statisticsto this diagram, Dr. F. Richard Yu and Dr. Victor C. M. Leung, own the highest number of published papers among other authors. Most of the published papers by these authors are also related to resource management and offloading topics given that these two areas are among the hottest research areas in the field of edge computing. However, some other authors such as Dr. Chunlin Li, Dr. Mohsen Guizani, and Dr. Tian Wang have focused more specially on security and privacy topics.

\begin{table*}[h]
\begin{center}
	\caption{Sample Publications of Active Researchers per Topic }\label{tbl3}
	\begin{tabular*}{\textwidth}{@{\extracolsep{\fill}} lcccccccc @{\extracolsep{\fill}}}
			\toprule
			Authors & CO & RM & NM & Ar. & S\&P & DM & Ec. & SUM  \\
			\midrule
			F. Richard Yu & 7 & 13 & 5 & 2 & 1 & 6 & 1 & 17 \\
			Victor C. M. Leung & 8 & 10 & 4 & 4 & 1 & 5 & - & 17 \\
			Yan Zhang & 9 & 11 & 2 & 5 & 3 & 2 & 2 & 16\\
			\bottomrule
		\end{tabular*}	

	\end{center}
\end{table*}

\begin{figure}[b]
	\centering
	\includegraphics[scale = 0.51]{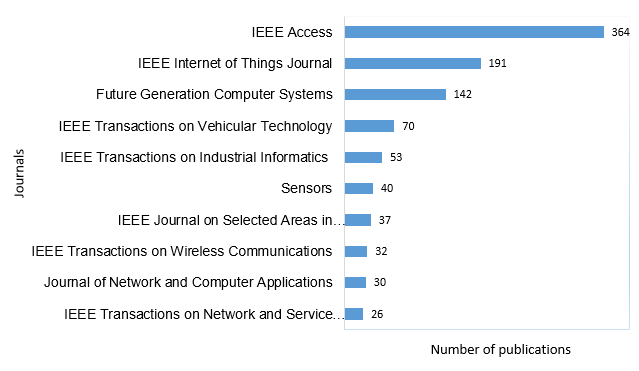}	\caption{Active Journals in The Field of Edge Computing}
	\label{fig:5}
\end{figure}

In order to identify active journals in the field of edge computing, similar to the process performed for active researchers, the number of papers published in each journal was counted. Next, the top ten journals with the highest number of relevant publications were selected as active journals. To this purpose, the information of the journal column in the [SuppFile]\_(E2,T2) is used. Information on the number of studies published by other journals per topic is also available in the [SupFile]\_(E4,T5). Figure\ref{fig:5} shows the active journals in edge computing with the number of papers published in each one. As illustrated in this figure, "IEEE Access", "IEEE Internet of Things Journal", and "Future Generation Computer Systems" have the largest number of published papers. Some reasons can be mention for the popularity of these journals. "IEEE Access Journal" is a multidisciplinary journal that covers all IEEE fields. The journal also expressly guarantees a review and publication time of 4 to 6 weeks. This feature can be a great interest as the papers on edge computing are being published with high speed. After "IEEE Access", "IEEE Internet of Things Journal" has the largest number of papers. The main area of this journal covers all aspects related to the Internet of Things including IoT system architecture, IoT communication, networking protocols, and applications such as smart cities, smart environments, smart homes, which are closely related to edge computing. Since no specific journal has yet been established to cover edge computing, this journal can be considered as one of the most specialized journals in the field of edge computing. The third journal is "Future Generation Computer Systems", which is one of the oldest and popular journals in computing, covering a wide range of computer systems.

\begin{figure*}[t]
	\centering
	\includegraphics[width = \textwidth]{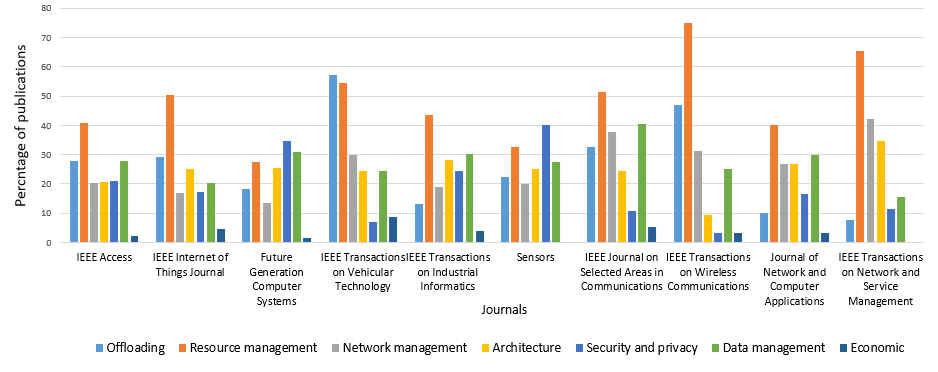}
	\caption{Percntage of Publications in Each Topic for The First Ten Journals}
	\label{fig:6}
\end{figure*}

Another analysis in this section is the percentage of papers published in journals per topic. Given the thematic diversity of the studies on edge computing, analyzing this information can also help the searching process in a particular sub-topic. Figure\ref{fig:6} shows the percentage of papers published in various topics for each of the pioneer journals. In this figure, the horizontal axis represents the journals, and the vertical columns represent the percentage of each topic publications in the corresponding journal. Because some papers may be categorized into more than one topic, the sum of the percentages in some journals may be more than one hundred. According to this chart, most of the papers published in “Future generating computing systems” and “Sensors” journals are related to security and privacy. The network management topic has higher percentage in “Journal of Network and Computer Applications”, “IEEE Journal on Selected Areas in Communications” and “IEEE Transactions on Wireless Communications” journals that focus more on computer networks. Resource management topics, which can be considered as the most challenging topic in this area, also account for the highest percentage of published papers in most of the journals.   

\begin{figure}[b]
	\centering
	\includegraphics[scale=0.42]{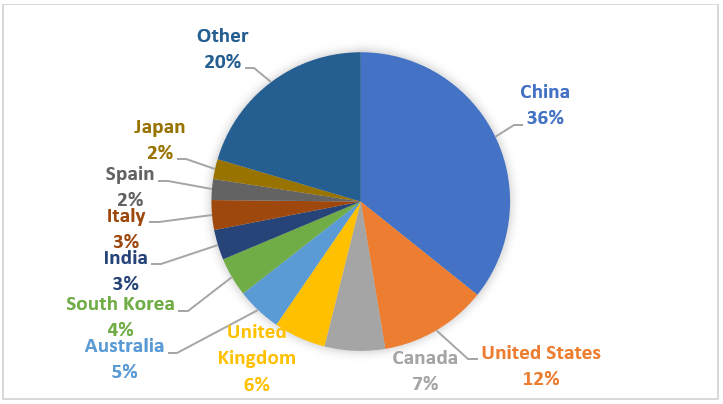}
	\caption{Geographical Distribution of Publications}
	\label{fig:7}
\end{figure}

The geographical distribution of research is another analysis that is discussed in this section. Familiarity with this information helps researchers to identify the pioneer countries in various fields of edge computing. The data in the [SupFile]\_(E1,T1) was used to produce the statistics. Table\ref{tbl4} shows some  of the contents of this file. Given that in each paper, different authors from different countries may have participated, so in this SMS, the criterion for determining the geographical location of authors is the countries mentioned in their affiliations. Figure\ref{fig:7} illustrates the percentage of participation of different countries in the reviewed studies. In this figure, the top ten countries with the highest number of papers are shown. The countries with low frequency are categorized in the ‘Other’ category. Information on the number of papers published from other countries is available in the [SuppFile]\_(E4,T7) file.

\begin{table*}[h]
\begin{center}
	\caption{Sample Affiliations}\label{tbl4}
	\begin{tabular*}{\textwidth}{@{\extracolsep{\fill}} c p{2cm} p{8cm} c c @{\extracolsep{\fill}} }
			\toprule
			PID & Authors & Affiliations & JID & Year \\
			\midrule
			14 & Changsheng You;
			Kaibin Huang;
			Hyukjin Chae;
			Byoung-Hoon Kim;
			 & Changsheng You, Department of Electrical and Electronic Engineering, The University of Hong Kong, Hong Kong;
			 Kaibin Huang, Department of Electrical and Electronic Engineering, The University of Hong Kong, Hong Kong;
			 Hyukjin Chae, LG Electronics, Seoul, South Korea;
			 Byoung-Hoon Kim, LG Electronics, Seoul, South Korea;
			  & 28 & 2016\\
			\bottomrule
		\end{tabular*}	

	\end{center}
\end{table*}

As shown in Figure\ref{fig:7}, China accounts for the bulk of published papers in edge computing (about one-third). There are several reasons for this observation. One reason is the existence of densely populated cities in this country, and the popularity of using smartphone among the Chinese people. Up to the end of 2018, China, with 850 million smart mobile users, had the largest number of mobile users among different countries, and this number is more than double the number of mobile users in the second country (India) \cite{r15}. This fact increases the need for new technologies such as 5G, and ultra-dense networks to cover users' access to the data network. Also, the need for edge computing to improve the user’s QoS is increasing. The Chinese, on the other hand, are pioneers in using technologies such as IoT, smart homes, and mobile applications based on artificial intelligence\cite{r16}. Various technologies including semantic analysis, speech and image recognition, have been rapidly deployed in the Chinese smart phone market. Ownership of various smart home-related products in China is also much higher than the global average\cite{r16}. Since the operation of these technologies and applications requires fast processing with low communication delays on edge resources, edge computing has received much more attention from Chinese researchers. After China, the United States, Canada, Australia, South Korea, the United Kingdom, Italy, India, Spain, and Japan have the highest number of papers, which are often the most high-tech countries in the SCImago Country Rank \cite{r17}.

\begin{figure}[b]
	\centering
	\includegraphics[scale=0.35]{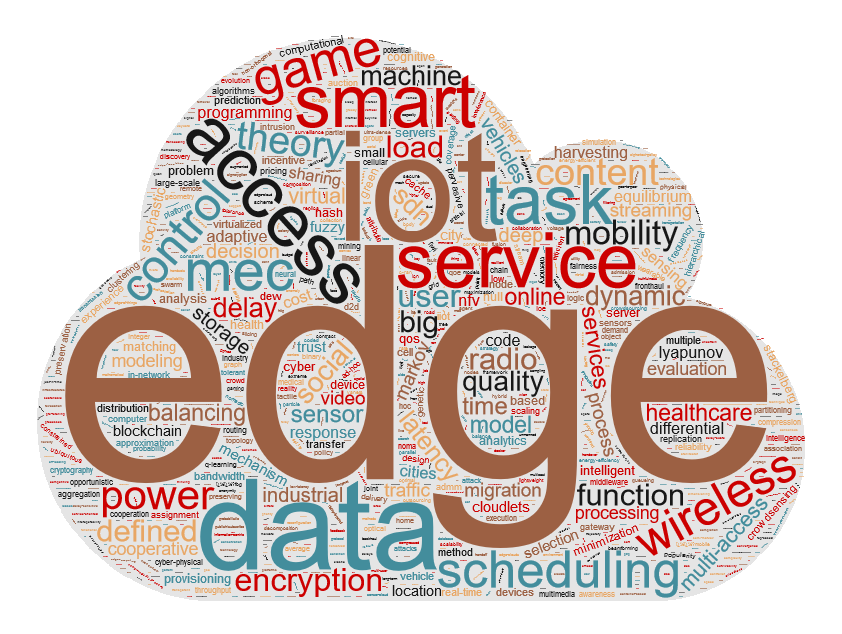}
	\caption{Main Keywords in the Field of Edge Computing}
	\label{fig:13}
\end{figure}

The final analysis in this section is about the keywords and common terms which are gathered from reviewed papers. Familiarity with keywords and common terms gives researchers a better perspective on the subject and helps them for better searching. For this analysis, the observed keywords in the keywords section of all papers are collected. For some papers that do not have a keyword section, some keywords related to the content of the paper have been proposed by an expert. In Figure\ref{fig:13}, a cloud diagram of the keywords with the highest repetition is shown which represents the most common terms in this field based on the reviewed papers. The larger the word, the more repetitions it has.

\subsection{What are the core research topics in the field of edge computing?}

In this section, from the perspective of reviewed papers, existing research scopes in the field of edge computing are discussed. According to Figure 1, these scopes are organized as topics and sub-topics in the built research tree. If the subject of some studies overlaps, they are categorized into more than one topic. There is also a sub-topic named miscellaneous, which includes miscellaneous fields that do not fit into any of the sub-topics. As mentioned earlier, in order to properly cover the topics of the research tree some other review papers have been examined. We have limited our SMS to journal articles and we discuss the effects and threats of this limitation on the article in Section 3.1.4 of Appendix A. However, these review papers cover conference papers and our obtained research tree shows that the topics are covered appropriately. Besides, to properly extract the topics and sub-topics, the covered topics (at least one topic) for each included article were selected using the title, abstract, and full text of the article. The complete information about these topics is reported in [SupFile]\_(E2,T4). In the following, each topic, along with its sub-topics, is explained.

\subsubsection{Computation Offloading}
This research topic includes papers whose innovation is precisely in the field of offloading. This topic has received the second most attention from the research community after resource management topic among the reviewed papers. MDs often face severe limitations on CPU, memory, and battery life. Offloading is one of the critical parts of edge computing, allowing users to run heavy applications on MDs through edge resources. The goal of offloading is mainly to speedup task execution, power saving, or both of them. The destination of offloading in various applications is the remote cloud resources, edge resources, or a combination of them (collaborative offloading). Using cloud and edge resources require network communication. Therefore, in the case of network instability, using collaborative offloading is a good option \cite{r18}. It should be noted that papers that only use the cloud resources as a destination (like \cite{r19}) do not fall within the scope of this SMS. These papers are generally in the field of MCC (Mobile Cloud Computing). In the following, the offloading research sub-topics are explained.

\paragraph{Decision Making}
This sub-topic includes those studies in which a user has different choices for offloading and therefore has to make decision. The decision-making problem, which plays an essential role in balancing the user's benefits and reducing the offloading overhead, has been expressed and resolved in various forms in the studies. In some studies, the issue of deciding between the three options of local execution, execution on limited resources available around the user (cloudlets or edge devices), or execution on unlimited remote cloud resources with high access latency has been raised \cite{r20,r21}. If it is possible to connect to other users' devices through the MANET network, it can also be considered as another offloading option \cite{r22,r23}. The issue of decision-making can also be raised about choosing from multiple resources of the same provider that are available close to the user. This type of decision-making is more common in ultra-dense networks where a user may have access to multiple macro or small cell base stations at the same time \cite{r24,r25}.

\paragraph{Mobility management}
Mobile edge computing is built based on cellular networks, and Wi-Fi access points, so mobility is an inherent feature, and an important factor in distributed cloud computing, and edge computing. User mobility can cause failures in offloaded requests. Therefore, it is necessary to design a suitable mechanism for mobility management to improve the QoS in edge computing applications. In the reviewed papers, various solutions to overcome the mobility issue are presented. The most important of these solutions are the methods of predicting mobility and user path \cite{r26,r27}, modeling the user mobility using Markov chain \cite{r28}, and movement pattern analysis \cite{r29}. User mobility and limited coverage of edge servers may lead to reduction in performance, service continuity and consequently, QoS violations. In this case, service migration is considered as a suitable solution, which will be mentioned in the topic of resource management.

\paragraph{Partitioning}
Another branch of research in the computation offloading topic is partitioning. Partitioning means splitting computation tasks or data to execute separately on different devices. This allows for parallelization, and helps run applications more efficiently. A critical issue in partitioning is the type of computational tasks. In general, two common models of binary tasks and partial tasks are considered in the researches \cite{r30}. Binary tasks have a high degree of integration, and are not decomposable; hence they must be run entirely on one resource. On the other hand, partial tasks can be broken down into smaller tasks, and can be executed in parallel. Partitioning of a task can be performed by breaking the code, and executing independent processes, and components in different resources \cite{r31} or breaking input data into different sections \cite{r32}. Partitioning can be done based on various criteria such as energy \cite{r33,r34}or resource efficiency \cite{r31}.

\paragraph{Computation Migration}
Computation migration involves the dynamic transfer of computing (in whole or in part) from UEs to near or remote sources. Some examples of these methods are the system VM method in cloudlet \cite{r35} and ThinkAir \cite{r36}, application-level virtualization in MAUI \cite{r37}, and imitation of ISA commands in \cite{r38}. In the method proposed in \cite{r35}, the user application in the form of a VM is migrated from the UEs to a resource at the edge. This method includes the entire VM with the program, which has more overhead than other methods \cite{r38}. In MUAI, application-level virtualization has been used to reduce overhead. In these methods, some parts of the code, such as functions or classes that need to be offloaded, are specified by annotation and offload using application-level virtualization like dalvik VM in android \cite{r39}. In the ISA imitation method (process migration), the amount of data to be offloaded is very less as compared to VM and application-level virtualization. Due to the difference between the architecture of mobile devices (usually with ARM) and the architecture of servers (generally Intel), Instruction Set Architecture (ISA) emulation is required while offloading process state between heterogeneous processors \cite{r38}.

\subsubsection{Resource management}
Resource management is almost the most critical issue in edge computing. Therefore, in the reviewed papers, it has attracted the most attention from the research community. Resources in edge computing can be divided into three categories: computational resources, storage resources, and network resources. Unlike cloud resources, these resources are available to users in a heterogeneous, limited, and distributed manner. These resources are also more dynamic than cloud resources, and in some cases, are distributed competitively among users \cite{r40}. Resource management in edge computing refers to a set of control processes for allocation and retaking resources to user tasks or requests according to various criteria such as latency, cost and energy. The most important research subfields in this topic are resource allocation and scheduling, migration and placemen, load balancing, resource estimating and utilization, and resource sharing, which are described below.

\paragraph{Resource allocation and scheduling}
Resource allocation and scheduling include mechanisms that allocate resources appropriately, and at the right time to specific tasks or requests. This allocation must be made under existing criteria and constraints. Improper resource allocation may lead to resource efficiency reduction, violating the energy constraints, or missing user deadlines. In the reviewed papers, scheduling methods have been used for several cases such as performing tasks execution on VMs \cite{r41} or containers on edge \cite{r42}, assigning user sessions to application samples \cite{r43}, running services on edge resources \cite{r44}, and network resource allocation \cite{r45}. 

\paragraph{Resource migration and placement}
A placement algorithm is responsible for selecting a proper place for things such as a service \cite{r46}, a VM \cite{r47}, and computing resource \cite{r48}. Live migration also refers to the transfer of a VM from one host to another without interrupting the VM service \cite{r49}. These two processes are usually done proactively (offline) and reactively (online) \cite{r2}. In a proactive method such as \cite{r50}, multiple versions of the service are already placed on different hosts. Reactive methods such as \cite{r48} are dependent on user mobility and due to this mobility, the most suitable edge node is selected for the migration or placement process. User mobility, path uncertainty and the dynamics and heterogeneity of resources at the edge of the network have caused the edge resource placement and migration to be more complex than cloud computing \cite{r51,r52,r53}.

\paragraph{Load balancing}
A load balancing mechanism tries to distribute the load evenly between the nodes, and prevent the load from concentrating on one or a few nodes. This directly affects the reduction of users' response time \cite{r54}. Because edge computing is based on a distributed platform and computational nodes are located in different geographical locations, the issue of load balancing is one of the main problems in this field. In the reviewed papers, load balancing problem is mostly solved by creating and distributing different versions of a program or a service on fog nodes \cite{r55} or different cloudlets \cite{r56,r57}, and then sending data and requests to these nodes based on the amount of load.

\paragraph{Resource sharing}
Resource sharing is a solution to overcome resource constraints at the edge of network, UE's limited resources. Generally, resource sharing deals with three issues: 1- the lack of a specific type of resource in the UEs to perform a task, 2- the lack of sufficient resources in the UEs, and 3- the use of other resources to speed up the completion of a job \cite{r58}. Resource sharing can be done between UEs through technologies such as D2D introduced by 5G, Mobile ad-hoc, Wi-Fi, and Bluetooth, or by creating clusters between edge resources. In resource sharing between UEs, network, computing, or storage resources are shared for faster, more secure, and better execution. For example, to improve network reliability, edge devices can use not only their communication resources, but also the heterogeneous network resources of other UEs to meet the QoS requirement. Using the resource sharing technique at the edge can not only increase resource efficiency, but also make it possible to run heavy applications \cite{r59}.

\paragraph{Resource estimation }
This sub-topic reviews papers related to resource estimation, one of the first steps in resource management. Resource estimation is a process that tries to predict the number of resources required to complete a task or overcome a computational load. Resource estimation is also an important process for dealing with fluctuations of resources requisition, and ensuring QoS \cite{r60}. Proper resource estimation brings efficiency, and fairness in resource management \cite{r61}. 

\paragraph{Resource utilization}
Resource utilization refers to optimizing resource usage based on relevant criteria such as time, energy, and cost. In most studies, two or more of these criteria are considered simultaneously. Resource usage optimization is often done by modeling the resource provisioning and consumption process with common optimization methods. The most widely used methods are linear programming \cite{r62}, PSO \cite{r63}, Lyapunov \cite{r64}, game theory\cite{r65}, heuristic \cite{r66}, and metaheuristic \cite{r41}.

\subsubsection{Architecture}
A portion of the reviewed papers has provided an architecture for performing distributed computations on edge resources, or has added capabilities to common edge architectures. Therefore, architecture is considered as a separate research topic. The purpose of the papers presented in this topic is to determine the architecture components, and interaction between them to provide new capabilities or facilitate the implementation of specific applications. The researches of this topic are divided into two sub-topics: Application-specific architecture and General-purpose architecture, which will be explained in the following. 

\paragraph{Application-specific}
This sub-topic includes researches that provides an architecture for running a specific application in edge computing environment. Since different applications have different features and requirements in terms of resource supply or network traffic, in some researches a series of modification at the infrastructure or the topology of elements in the edge architecture have been done to support a special application requirement. Among the researches in this sub-topic, we can mention the presentation of architectures for healthcare applications \cite{r67}, crowdsensing applications \cite{r68}, and applications of automated monitoring systems \cite{r69}.

\paragraph{General-purpose}
This research sub-topic is about designing an all-purpose architecture that provides a basic configuration for distributed edge computing. In these studies, the goal is to create a novel or modified architecture according to new criteria, or to add new capabilities to previously known architectures. Examples of these studies include providing an architecture for agreement and consensus between nodes on a particular community \cite{r70}, trust evaluation \cite{r71}, reducing energy consumption \cite{r72}, and  providing an architecture based on blockchain for security \cite{r73}.

\subsubsection{Network management}
One of the most important foundations in which is found in reviewed papers is the network management issue. This research field focuses on optimizing parameters related to network infrastructure, and applying technologies and capabilities of new generations of cellular networks to edge computing paradigm. Also, this topic covers all aspects of modifying the current standard network architectures to adapt to the edge computing capabilities. In the reviewed papers, various researches in the field of network management have been observed. Several research sub-topics in this category have been classified, which are often based on standard network technologies. In the following, each of them is explained.

\paragraph{Access Control}
In area of communication, there are two well-known concepts: network access architecture and multiple access schema. Conceptually, the access network such as the RAN (Radio Access Network) connects a mobile device to its core network (CN) by residing among themselves. On the other hand, a multiple access schema (also called channel access method) such as NOMA (non-orthogonal multiple access) uses multiplexing to share a communications channel or physical communications medium between multiple users \cite{r74}. In reviewing the studies, the literature on these concepts was examined in a single topic and described below. 

\textit{Access network architectures: }Radio Access Network (RAN) is a famous architecture in mobile and cellular networks that have evolved with the advancement of generations of mobile communications. Conceptually, it connects a device such as MD or a computer to its core network (CN) by residing between them. RAN network is consists of a set of base stations (BS). Each BS, depending on its transmission power, covers a specific area and is separated into a Remote Radio Unit (RRU) for radio functionalities and Base Bound Unit (BBU) for baseband processing and channel processing. The data transmission between RRH and BBU are done with Common Public Radio Interface (CPRI) \cite{r2} Cloud-RAN (C-RAN), is a version of RAN in that has the potential to handle as many BS as the network needs using the concept of virtualization. In C-RAN, the BBUs is virtualized and shared among operators in centralized BBU pool \cite{r75}. In Fog RAN (F-RAN) architecture, the central computing in C-RAN is transferred to the edge of network. In this architecture, a fog node has the ability to cache contents and provide computation capabilities. This concept was first introduced by Cisco to exploit local signal processing, and computing, collaborative resource management, and distributed storage and caching at the edge of network \cite{r76}. Since then, a part of the research in the field of edge computing has focused on combining the MEC with this technology. Research in this area has often focused on designing new F-RAN-based architectures for resource management \cite{r77,r78}, adding mobility management mechanisms \cite{r79}, or providing methods to increase resource efficiency \cite{r80,r81}.

\textit{Multiple access schemas: }Two well-known Multiple Aaccess (MA) schema in wireless communication systems are NOMA and OMA. The main difference between OMA and NOMA is that in OMA bandwidth is divided between UEs based on frequency (such as FMDA), time (TDMA), or code (CDMA). But in NOMA, the bandwidth will be divided fairly between the UEs \cite{r82}. For example, With OMA, connecting thousands of IoT devices, such as vehicles in vehicular ad hoc networks for intelligent transportation, requires thousands of bandwidth channels; however, NOMA can serve these devices in a single channel use. An important phenomenon in NOMA networks is that some users with poor channel conditions will experience low data rates.  NOMA is one of the most promising radio access techniques in next-generation wireless communications, i.e., 5G \cite{r83}. Cognitive radio (CR) is a form of wireless communication in which a transceiver can intelligently detect which communication channels are in use and which are not, and instantly move into vacant channels while avoiding occupied ones. NOMA is special case of CR \cite{r84}.

\paragraph{Network abstraction and Orchestration}
This sub-topic is related to researches that take advantage of Software-Defined Networking (SDN) technology in edge computing environment. SDN is a centralized network configuration architecture that enables centralized programming, and control of network traffic. In this network configuration method, the control plane is separated from the data plane, and is managed by a central module with a global view of the traffic, and the overall state of the network. Network Function Virtualization (NFV) is a technology that is usually used as a complement to SDN. NFV refers to the process of virtualizing the network equipment functions, in which networking functions are performed on computing resources such as VMs or containers located in the cloud or at the edge of the network. The use of these two technologies rapidly change the development of network functions and the evolution of network-based architectures. They brought valuable benefits such as cost reduction, increasing network flexibility and scalability, and reducing the time-to-mark time of new applications, and services \cite{r85}. 

The decentralized feature of edge computing environment is one of the inherent and main issue that causes several problems for computing, storage, security, and traffic control. The use of centralized control, and decision-making capabilities, and programmability in SDN has led this technology to be widely used in edge computing researches. These solutions include use of SDN in resource management \cite{r85}, cloudlet management \cite{r86}, traffic control \cite{r87,r88}, and mobility management \cite{r89}.

\paragraph{Traffic management and engineering}
This research sub-topic is related to traffic modelling optimization, and engineering in edge computing, which is one of the vital processes in networks. One of the prominent features of edge computing is heavy traffic at the edge of network due to the high volume of user requests for different services. Therefore, a portion of the researches in the field of network in edge computing has focused on traffic management and engineering. Traffic engineering is a mechanism for optimizing network resources, and providing the requirement of services by allocating bandwidth and selecting the route of traffic in the network \cite{r90}. In the reviewed papers, the video streaming applications were the frequent applications in the field of traffic management, and engineering \cite{r90,r91} as the traffic related to video streaming services account for the largest part of network traffic \cite{r92}.

\paragraph{Slicing and Overlaying}
Network slicing is a technology that has been proposed to meet various applications, and different business models in 5G. This technology has been highly regarded in the academy, and industry. The concept of network slicing is the division of physical network infrastructure entities into isolated logical network components with appropriate functions, which is done to meet the different application requirements. Each of these network components is used for a specific purpose or service \cite{r93}. As the requirements of edge-based computing services are highly diverse, network slicing is a flexible, and promising solution to meet the needs of various services \cite{r94}. In the reviewed papers, the network slicing technique has been used for various purposes such as separation of MEC services from traditional services \cite{r95}, division of computational resources by considering the energy criterion for various services requested by users \cite{r96}, and providing application-based QoS \cite{r97}.

\subsubsection{Security \& privacy}
Like any other infrastructure, edge computing is not devoid of aggressive and hostile agents. Edge computing is based on a distributed and almost unreliable platform. Therefore, security in edge computing has been considered at various levels in the literature. The most critical security issues in this field are the design of access control mechanisms, privacy-preserving, trust management, and attack prevention, which are explained below.

\paragraph{Access control}
The purpose of access control is to design a mechanism for monitoring and ensuring that data is accessible only to authorized people who have access permission. Access control is an important security feature, especially for the applications that are related to user data, such as storage applications. Attribute-based encryption is one of the well-known techniques for providing access control which has a significant encryption/decryption computation overhead. Some researches focus on reducing this overhead through outsourcing the encryption/decryption computation to edge servers \cite{r98,r99,r100,r101}. The blockchain technology, which has recently attracted the attention of researchers in a wide range of industries, is also considered in the researches of this area \cite{r73,r101} are some examples of these works.

\paragraph{Attack detection}
In edge computing, various attacks can be imagined in different levels of the network or application that can cause damage to varying degrees. The purpose of this research sub-topic is mainly to prevent or identify these attacks. The "Man in the Middle Attack" \cite{r102}, "Selective Forwarding Attack" \cite{r103}, and "Data Injection Attack" \cite{r104} are some examples of common attacks in this field.

\paragraph{Trust evaluation}
Trust value measurement among the components of a community is one of the security features that is considered in detecting attacks and securing systems. Determining the degree between pairs has particular importance in edge computing due to the existence of low-reliable heterogeneous edge nodes. Therefore, several studies have been conducted in the field of trust in edge computing \cite{r105,r106,r107}. Among the applications considered in this field, trust in social networks has always been considered. In these networks, trust can be verified by a user, a service provider, or by observing a connection between two entities \cite{r108}.

\paragraph{Privacy}
Most of the edge computing applications are inherently based on content uploading. Therefore, the probability of threatening users' privacy is high. This issue has received more attention in personal data-driven applications like crowdsensing \cite{r107,r109} and healthcare applications  \cite{r23,r110}. In Healthcare applications the health-related data of the users which are measured by wearable sensors are sent to the nearby edge servers to be analyzed. In Crowdsensing applications, the sensed data of volunteer user devices are collected to be used in a specific application. In both of these scenarios the private user data may be accessible by the application providers. Another common private user data that can be violated through edge computing applications is the user location which is used for navigation or map-based applications. The user location privacy preserving has also received more attention in researches \cite{r111,r112}.

\subsubsection{Data management}
With the advent and expansion of IoT and the dramatic increase of smart devices that generate data, data management has become one of the most critical issues in the field of edge computing \cite{r113}. Therefore, in recent years, much attention has been paid to providing efficient methods for data management. These methods include collecting, storing, caching and processing data. So far, these operations have been done in the remote cloud with good quality. But in recent years, with the increase in data volume, and bandwidth limitations of backbone network, this will not be possible. Edge computing is a right solution for this problem, in, which data is sent to the nearest edge equipment instead of the remote cloud. After reviewing the papers in this field, several research sub-topics were identified, which we will explain below.

\paragraph{Caching }
This sub-topic includes researches related to network edge caching. Caching is a well-known technique to speed up data access, which reduces the load of the backbone networks, and improves the quality of the user experience in the edge. In this technique, frequently used data traffic such as news, weather, videos, and popular contents are cached at the edge of the network.  As a result, the load of the backbone network and the content access delay is reduced for users. Besides, the use of this technique can be significantly effective in increasing performance \cite{r114}. One of the most common issues in caching is predicting the popularity of content that should be cached \cite{r115}. In the reviewed papers, various methods have been used to solve this problem, such as learning methods \cite{r116}, location utilization \cite{r117}, and methods based on social networks \cite{r118,r119}. Caching can improve the performance of computation offloading in wireless networks. The goal is to cache the result of computation and avoid to process the same task \cite{r120}.

\paragraph{Data analysis}
Many services, and applications related to the IoT field, including smart city, smart transportation, healthcare etc. are working based on data. Data collected from UEs or sensors must be sent to data centers for analysis, and knowledge extraction. One of the typical applications in edge computing is performing all or some parts of the data analysis operations at the edge of network. This reduces the amount of data sent to data centers, and speeds up the users request response time. Data analysis is performed for different purposes at the edge from simple analyzes such as data aggregation and preprocessing, matchmaking, and filtering \cite{r121,r122,r123} to more complex data mining applications \cite{r124}. The use of big data techniques such as stream processing, and their adaptation to edge infrastructure has also been considered in this area \cite{r125,r126}.

\paragraph{Data distribution}
With the advent of new applications, storing data in the cloud is practically inefficient due to high traffic between the user and the cloud, high latency, and cost \cite{r127}. Due to the limited storage resources available at the edge compared to the cloud, the distribution of data between partner devices at the edge can solve this problem. Node mobility and determination of the geographical location of data are significant challenges that should be considered in this field. In \cite{r128} the authors present an architecture for distributing large volumes of data across automotive networks for content sharing. In most CDN applications, including \cite{r129,r130}, content placement is a challenge that deals with distributing content appropriately between resources and devices at the edge \cite{r129}.

\paragraph{Data dissemination}
Data dissemination refers to the mechanisms that determine the optimal path for data transfer. Data dissemination can occur between UEs, between nodes at the edge of the network, and between edge nodes and remote cloud. In \cite{r131}, the a k-means clustering algorithm was used to transmit data thorough longer distances but with less energy consumption in a LPWAN network. In this research, devices were placed in different clusters based on traffic priorities. In \cite{r132} to solve the problem of high data costs and reduce the traffic between fog and cloud nodes, a combined data propagation framework using SDN and DTN is presented. In this framework the control plane is resided in the cloud, and the data plane is resided at the fog nodes.

\paragraph{Data replication}
Data replication is one of the most common techniques in cloud computing, in which a set of data is stored in different data centers with different geographical locations to improve availability and reliability \cite{r133}. This method is also used in edge computing. Applying this technique in edge computing can improve the access speed, QoS, and bandwidth consumption \cite{r134}. However, replication mechanisms designed for the cloud may not be appropriate for edge servers due to limited resources and varied traffic in edge RANs. Thus, new replication methods are needed for the edge computing paradigm \cite{r135}.

\subsubsection{Economics}
This topic includes researches in, which economic aspects are considered along with other criteria. By comparing the number of papers published in this section, it can be seen that this field of research has the lowest number of papers and many research opportunities in this field can be imagined. In the papers reviewed in this field, economic theories have been used to increase resource productivity, maximize profits, and creating a fair environment between providers and customers.

The main economic research problem in reviewed papers is pricing that deals with issues such as designing mechanisms for determining the price of resources, and setting penalties for violation of the guaranteed QoS. In general, pricing is a strategic issue between the supplier and the buyer. The supplier seeks to increase resource productivity and profit, and the buyer seeks to minimize the cost as long as the time and energy constraints are met. Therefore, in the reviewed papers, matching and game theory techniques have been widely used. The most commonly used game theory techniques are Stackelberg game \cite{r136,r137}, coalition game \cite{r138,r139}, and market game \cite{r140,r141}.

One of the well-known and widely used pricing techniques in the edge computing environment is auction-based pricing. The limitation of edge server resources and the fluctuation of user requests, which may create a competitive environment especially at peak time. Auction is one of the most common ways to allocate resources efficiently and fairly in such cases. In an auction, there are two main agents: the seller, and the buyers. Typically, in edge computing, buyer agents are mobile users, and seller agents are service providers or cloudlets. In some cases, the mobile users themselves can also appear in the role of the seller \cite{r142}. A buyer agent sends a request with a bid price when he/she need to get a resource. Next, the sellers select the buyers based on the offered bid prices, and provide the resources to the winner buyers. Items auctioned in edge computing can be merely processing resources \cite{r143,r144} or processing, and communication resources together \cite{r145,r146}.

\subsection{The distribution of applications in each research topics}
In this section, we discuss the common applications in the field of edge compuing and the reasons why these applications are suitable for running on the edge platform. Furthermore, the distribution of applications in the topics of edge computing has been investigated. These applications are extracted from papers and are listed in Table\ref{tbl5}. It should be noted that in this SMS, papers that introduce their innovation in general and do not limit it to a specific area, are reported in the General category. Also, papers with very few repetitions that cannot be added to any of the existing classifications of applications, fall into the Other category. The Other and General categories have been neglected in the following comparison between applications, in order to emphasize on mostly used applications in more depth. 

\subsubsection{IoT Applications}
IoT refers to connecting enormous number of physical devices to the internet  to sense, collect and share data. In recent years, IoT as a growing technology has received much attention and has been used in many applications. However, the increasing volume of data collected by sensors and other devices used in this environment has made it difficult to process data on IoT devices, which typically have low memory and CPU power. Initially, cloud computing was recommended as a solution to this problem, but it also had some drawbacks. Bandwidth and latency issues in sending and receiving data are the most critical barriers to cloud computing in IoT applications \cite{r2}. Therefore, edge computing has been considering as a suitable solution. Edge computing can overcome the problems resulting from the network and its latency by sending data to computing resources at the edge of network. As can be seen in Table\ref{tbl5} papers in the IoT category have the highest statistics which shows the high compatibility of the iot applications with edge computing. These paper are mostly classified into the resource management, security, architecture, and computation offloading topics.

\begin{table*}[h]
	    \begin{center}
	        \begin{minipage}{\textwidth}
	        \caption{The Distribution of Applications per Topic.}\label{tbl5}
		\begin{tabular*}{\textwidth}{@{\extracolsep{\fill}} p{3.5cm} c c c c c c c @{\extracolsep{\fill}} }
			
			\toprule
			\diagbox{Application}{Topics} & \begin{sideways}Co\end{sideways}       & \begin{sideways}RM\end{sideways}       & \begin{sideways}DM\end{sideways}       & \begin{sideways}Ar.\end{sideways}      & \begin{sideways}NM\end{sideways}       & \begin{sideways}S\&P\end{sideways} & \begin{sideways}Ec.\end{sideways}      \\
			\midrule
        General	& 185	&	311	&	114	&	99	&	140	&	92	&	28 \\	
IoT	&	76	&	132	&	60	&	75	&	55	&	75	&	9\\	
Smart Community\footnotemark[1]	&	9	&	30	&	22	&	36	&	13	&	20	&	1 \\
Vehicular	& 28	& 50	& 32	& 38	& 39	& 24	& 3	\\
Healthcare	& 8	& 14	& 30	& 28	& 8	& 19	& 0	\\
Streaming	& 10	& 18	& 19	& 7	& 8	& 3	& 0	\\
Industria	& l4	& 12	& 9	& 12	& 6	& 4	& 0	\\
Crowdsensing	& 2	& 5	& 3	& 8	& 2	& 10	& 0	\\
AR/VR	& 2	& 9	& 3	& 4	& 3	& 0	& 0	\\
Other	& 10	& 15	& 25	& 16	& 11	& 12	& 1	\\
		\end{tabular*}
		\footnotetext[1]{Smart Community (Including Smart City, Smart Home, Smart Grid, etc)}
		\end{minipage}
	    \end{center}
	\end{table*}

\subsubsection{Vehicular Applications}
Vehicular applications refers to applications that increase the quality of the driving experience.This application is also one of the most frequent applications in edge computing papers and researchers amaze vehicle drivers with useful applications in various fields such as traffic management \cite{r147}, parking reservation \cite{r148}, and more. Many vehicle applications, such as autonomous driving or traffic management, require specific processing, storage, and communication capabilities. Edge computing with features such as presence near the user, low latency, high mobility support, real-time communication, context-awareness support, and low development costs is an excellent solution to meet the requirements of vehicular applications \cite{r149}. Network stability challenges due to vehicles mobility as well as the fast response requirement of these applications are of the most important isses in these applications. Therefore the majority of the papers of this application fall into resource management, network management and architecture topics which are dealing with mentioned issues.

\subsubsection{Smart Community}
This category contains all applications that have used artificial intelligence techniques to increase their capabilities in the smart community. The majority of the papers of these applications fall in the architecture topic. Examples of such papers include, managing smart city applications \cite{r150}, connecting IoT tools to a smart hospital \cite{r151}, and providing an architecture for big data management with a real-time manner for smart transportation applications \cite{r152}. All these architectures are edge/fog-based.

\subsubsection{Healthcare}
The healthcare applications are monitoring systems that collect patient’s data by himself or through wearable sensors to checks the patient's daily activities, eating habits, sleeping patterns, giving helpful recommendations for a healthy lifestyle or detecting medical emergencies \cite{r153}. The massive data generation due to continuously sensing and the fast response requirement especially for emergency situation caused this application to be widely explored in edge computing paper. Since these applications work based on the analysis of patient’s personal health data, the privacy is a very important issue. So data management, architecture and security and privacy are the prominent topics for this application.

\subsubsection{Streaming}
This category refers to the applications in which a continuous and live stream of data must be processed continuously. One of the most common examples of these applications are multimedia streaming applications. The real-time processing requarement is an important feature of these applications and high latency data transmission during execution, severely degrade the user quality of service. Edge computing can be used to increase the quality of multimedia services by bringing processing and storage resources closer to user devices. Resource management and bandwidth allocation to control the latency and jitter \cite{r154,r155} as well as managing cached data \cite{r156,r157} are of the most important issues in this category of applications, so the articles related to these applications are mostly categorized in the resource management and data management topics.

\subsubsection{Industrial}
Industry applications are another category of applications that has been targeted by edge computing researchers. After transition to the fourth revolution of manufacturing which is called industry 4.0, issues such as communication between industrial equipment and data analysis were raised to increase automation in decision-making and reduce human involvement \cite{r158}. However, these applications require the rapid processing and analysis of large volumes of generated data \cite{r159}, which can be conveniently performed through edge computing.

\subsubsection{Crowdsensing}
Crowdsensing is a type of application in which a large scale task is distributed over a crowd of people who are voluntarily  participating via mobile devices. These taks include a wide range of applications which are mostly dealing with sencing data via mobile sensors, reporting a situation or doing a simple and small part of a big job. Many companies use crowdsensing to obtain large amounts of low-cost data for useing in their online services \cite{r160}. Due to the widespread acceptance of this application and the increase in the volume of sensed data traffic, edge computing has become one of the suitable options for implementing these applications. However, this model of application faces various threats due to its voluntary nature, including sabotage and sending false data and user privacy. Therefore, privacy and security is a major challenge in these works and a significant part of the articles in this application fall in the privacy and security topic. To address these challenges several reputation management and privacy preserving thechniques are presented in the edge computing literature \cite{r161,r162}.

\subsubsection{Ar/Vr}
Augmented and virtual reality are among  the  most  attractive  applications in recent years. Virtual reality detects and tags objects or adds some extra elements to a live video capturing by user smartphone and augmented reality gives the user the experience of interacting in a completely virtual world through VR headsets \cite{r163}. These applications requare fast data communication and analysis to provide user an immersive real-time interactions \cite{r164}, thus they have received a lot of attention in the edge computing reaserch community. These applications similart to streaming applications are very delay sensitive and increasing the quality of service through reducing latency is the main challenge of such applications which can be achived by choosing a pooper resource allocation, task placement and decision making policies \cite{r165,r166,r167}. This is the reason why the articles of these applications fall more in the resource managemen topic.

\subsection{The distribution of architecture in each topic}
\begin{figure}
	\centering
	\includegraphics[scale=0.51]{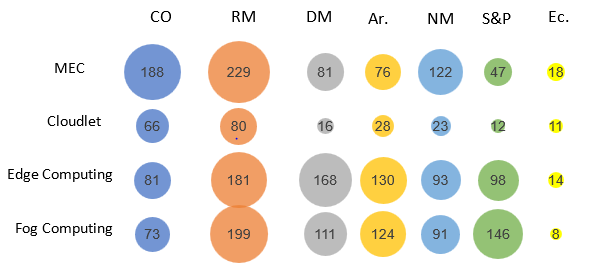}
	\caption{Architecture per Topics}
	\label{fig:8}
\end{figure}

In this section, we statistically analyze the edge architectures used in different topics in reviewed papers. Figure\ref{fig:8}, shows the frequency of utilization of different architectures by researchers. In this figure, the horizontal axis represents the topics, the vertical axis represents the architectures, and the volume of the bubbles represents the number of articles using the related architecture in the relevant topics. The main architecture in edge computing includes fog, MEC, and cloudlet. However, in some studies on the field of edge computing, none of these architectures are mentioned and only the word "edge computing" is used. In this paper, we have categorized these studies as general architecture and investigated them under the same heading as edge computing. 

In the reviewed papers, various architectures have been proposed to take advantage of close resources. The most well-known, and widely used architectures are fog \cite{r76}, cloudlet \cite{r35}, mobile edge computing \cite{r168} (also known as Multi-access edge computing), and mist computing \cite{r169}. Although these concepts have some similarities, and are sometimes used interchangeably in papers. But there are some differences between them.

Fog computing architecture uses devices around the user, from servers and computers to routers, and switches on the edge of the network, to process or pre-process before sending a request to the cloud. In fact, this architecture is a hierarchical architecture in, which all the processing capacities available in the user path to the remote cloud are used \cite{r4}. The cloudlet architecture includes a high-power computer or cluster of computers near the users \cite{r35}. Surrounding MDs can use the resources available in the cloudlets (often virtualized, and provided in the form of a VM) to increase their processing power. Cloudlet operators can be cloud service providers who intend to provide their services close to the user \cite{r4}. MEC is the development of mobile computing through edge computing which IT, and cloud computing capabilities are provided through RAN (Radio Access Network) in 5G and 4G \cite{r170}. In this architecture, edge computing services (processing, storage, and network) are provided through the infrastructure of telecommunication networks. This paradigm was further developed to cover a wider range of applications (beyond the specific tasks of mobile phones) under the name Multi-access Edge Computing \cite{r171}. Finally, the concept of Mist computing is a computational model for performing scattered computing at the most end nodes, i.e., the IoT devices themselves. Since this architecture is managed by the devices themselves, it has more complexities than other architectures \cite{r169}.

As shown in Figure\ref{fig:8}, the cloudlet architecture has significantly lower publications compared to the other three architectures. Cloudlet architecture is one of the first and most important models for increasing the speed and quality of mobile devices, which provides cloud services to users more quickly by bringing servers and computing resources closer to mobile users. But the emergence of newer and more general architectures caused the research community to pay attention to other architectures. For instance, fog computing can offer a more generic alternative and allows resources to be anywhere along with the edge devices to the remote cloud \cite{r2}. Furthermore, due to recent developments in the field of mobile networks, operators are focusing on MEC as the main edge server technology instead of Cloudlets \cite{r3}. There are also some meaningful differences in the number of papers of different architectures in different topics .The reasons of these observations are in the difference in the structure, features, and applications of each architecture. In the following, we discuss some of these features. 

In the offloading topic, a considerable number of researches have been done on the MEC architecture. Offloading is a well-known technique for reducing user device energy and increasing execution speed, so this technique has been widely used for mobile applications that face severe memory, processing power, and energy limitations. A significant portion of the publications in the offloading topic is about mobility management \cite{r27}, decision making \cite{r172}, and partitioning \cite{r173} for mobile applications, which are mostly conducted on cellular networks and MEC architecture. 

In the resource management topic, the number of publications in three main architectures has a high number as it is the most challenging topic in the edge computing. In the data management topic, the studies conducted on the MEC architecture is also less than the other two architectures. According to this study, MEC applications are mostly related to mobile applications and less used for data-intensive applications. One reason for this is the constraints on the location of the servers in the MEC architecture. The MEC architecture is placed on a RAN, in which the servers must be adjacent to the cellular network base stations, but in the other architectures, there is no such limitation \cite{r3}. So, in MEC architecture, it is not possible to bring servers closer to the data source. Another reason for the fog preference to MEC architecture in data management topic is the multi-layered architecture of fog and the presence of cloud support so that, fog nodes can be used to pre-process or filter data before sending it to the cloud for storage or analysis.

In the architecture topic, the MEC had fewer publications than the Fog and edge architectures. The publications categorized in this topic are related to researches that are trying to increase the efficiency of a particular application or improve a general parameter by making changes in standard architectures. The MEC has a more rigid and well-defined structure than other architectures due to its envisioned existence in a telecommunications infrastructure that is inherently regulated \cite{r3}. So, the MEC architecture has low flexibility for structural changes. But in other architectures, there is more flexibility to make changes and apply new architectural ideas. 

Finally, in security and privacy topic, fog architecture accounts for the most portion of the publications. In fog architecture, unlike MEC, the fog nodes are not owned by a provider and are provided by independent individuals \cite{r3}. Because fog nodes are usually deployed in some places with relatively weak protection, they may encounter various malicious attacks \cite{r174}. Furthermore, devices in the Fog are often deployed without strict monitoring and protection \cite{r175}. Therefore, security issues in this architecture have received more attention than other architectures. Another reasonable point in the Figure 8 shows that, in general,is that the number of publications on the economic topic is lower than on other the others. However, the two paradigms of MEC and edge are more welcomed than others. This indicates the particular need for the edge computing environment to design pricing and auction models. 

Paradigms such as osmotic computing, mist computing, application-centric computing were also observed in the papers, which do not have significant statistics. For example, Sharma et. al. utilized osmotic computing to propose a trust management framework for social networks applications\cite{r108}. Rahman et al. proposed integrating five-tier cloud, fog, and mist computing environments for IoT applications in healthcare and next-generation e-healthcare systems. Their proposed method aims at real-time handling and routing offline/batch data with high QoS and low end-to-end latency \cite{r176}.

\subsection{Techniques used in the field of edge computing}
\begin{figure*}
	\centering
	\includegraphics[width = \textwidth]{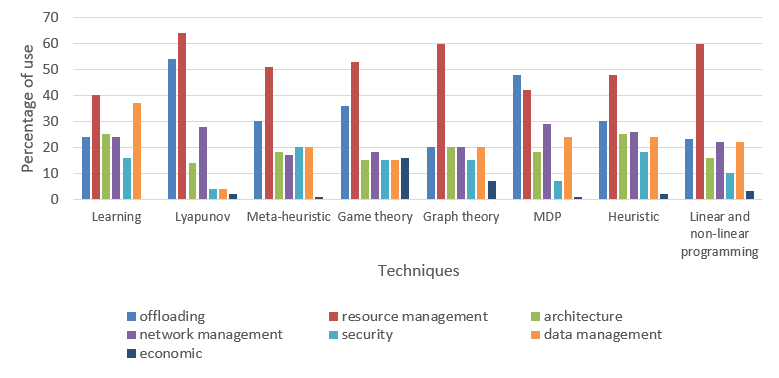}
	\caption{The Percentage of Technique Usage in Different Topics}
	\label{fig:9}
\end{figure*}

\begin{figure}[b]
	\centering
	\includegraphics[scale= 0.6]{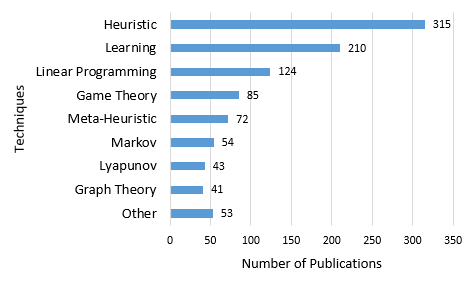}
	\caption{The Statistics of Techniques}
	\label{fig:10}
\end{figure}

With the advent of edge computing, new challenges have emerged in the development and management of distributed resources. Transferring storage and computation to the edge of the network increases the QoS and QOE, especially in latency-critical applications. However, it is difficult to use heterogeneous resources efficiently to meet the various needs of the users. To overcome these problems, various contributions have been presented in the literature in which different techniques have been used. Figure\ref{fig:10} shows the statistics of the most commonly used techniques in the field of edge computing. The percentage of techniques used in different topics are also shown in Figure\ref{fig:9}. Because some articles are categorized into more than one topic, the sum of the percentages may be more than one hundred. In the following, these techniques are explained with the focus on the reason why they are widely used in issues related to edge computing. 

\subsubsection{Heuristic techniques}
Heuristic techniques are a right solution for finding near-optimal values in problems where full optimization is difficult and time-consuming. The term heuristic refers to a method which, on the basis of experience or judgement, seems likely to yield a near-optimal solution to a problem \cite{r177}. These techniques begin with a well-defined mathematical representation of a problem \cite{r178}.  Since time is one of the most essential QoS criteria in edge computing applications, heuristic methods that find near-optimal answers in less time have been widely used by researchers in this field. The most important features of heuristic methods are easy implementation, the ability to show improvement in each iteration, fast production of results and robustness \cite{r179}. The most common use of heuristic techniques is in the resource management field. Many studies have tried to manage resources in such a way that energy \cite{r179,r180}, latency \cite{r181,r182}, cost \cite{r52,r182} and computational complexity \cite{r183} become optimized. As shown in Figure 9, out of 315 articles related to this technique, 30\% of the articles are fallen in the offloading topic, 48\% in the resource management topic, 25\% in architecture topic, 26\% in network topic, 18\% in security topic, 24\% in data management topic and 2\% in the economic topic. Heuristic techniques are general methods for solving many scientific problems, so this technique has a high percentage of usage in almost all topics especially in resource management in which many np-hard problems are mentioned in the articles.

\subsubsection{Linear and non-linear programming}
Linear programming is another way to solve optimization problems. This optimization method tries to solve complex problems by modeling them mathematically. In fact, objective functions and constrains are modeled by linear functions in this method \cite{r184}. In contrary, if one or more of the functions appearing in the problem statement are non-linear in their variable, the optimization problem called non-linear \cite{r185}.  

LP problems are part of a group of optimization problems called convex optimization. In this model, each problem is formulated as a set of goal(s), namely objective function(s), and constraint(s). A convex problem can be solved by applying the KKT condition or the lagransian method. Because in these problems the local optimal is the same as the global optimal, the problem can be solved and the global optimal can be easily obtained.

Some nonlinear problems, such as MINLP, may be modeled as non-convex problem. The easiest way to solve these problems is to turn them into a convex problem. For this purpose, some variables are considered as constants. The difficulty in solving these problems is due to the fact that in this category the local optimal may not be the same as the global optimal.

Convex and non-convex optimizations are centralized methods. In other words, in these models, there is a centralized agent that gathers all the necessary information and performs decision-making behalf of all. Therefore, these methods are not scalable and are only suitable for small scenarios. Also, Because due to static nature of convex and non-convex optimizations, they are unsuitable for dynamic applications. However, as mentioned before they can achieved a global or near global optimization.This technique has been used in 124 articles. 23\% of the articles are fallen in the offloading topic, 60\% in the resource management topic, 16\% in architecture topic, 22\% in network topic, 10\% in security topic, 22\% in data management topic and 3\% in the economic topic. As shown in Figure\ref{fig:9}, this technique has received much attention in resource management topic compared to other topics because it is a common method for solving a wide range of optimization problems in resource management. Some examples of these problems are balancing energy consumption and delay at the edge \cite{r186}, transporting, placing, storing, and processing huge amount of IoT data in an efficient and effective manner \cite{r187}, maximization the storage utilization while reducing service latency and improving energy savings \cite{r188} and optimization of tasks offloading over mobile edge computing environment \cite{r189}.

\subsubsection{Game theory technique}
Game theory techniques in IoT-based computing environments such as edge and fog computing help to solve the competition between different actors. The network is full of distributed resources that must be allocated equitably among IoT devices and users. On the other hand, the allocation of these resources should be done by considering the profit of the service providers. This is where game theory comes to the aid of researchers, and the variety of different games has led to it being widely utilized in papers after heuristic and learning techniques. 

Game theory techniques are distributed in nature, meaning that the decision-making is done locally by each user separately. Therefore, these methods are scalable and can be used for large and dynamic scenarios. The main drawback of Game theory techniques modeling is that they only reach the Nash equilibrium, which may not necessarily be optimal.

Among the 85 articles that have used game theory, 36\% of the articles are fallen in the offloading topic, 53\% in the resource management topic, 15\% in architecture topic, 18\% in network topic, 15\% in security topic, 15\% in data management topic and 16\% in the economic topic. One of the noticeable points in this statistic is the high percentage of economic articles compared to its value in other techniques. Most of the articles in the economic topic are related to pricing, which is a strategic problem. In these type of problems, each agent seeks to increase its benefit, and game techniques are a good solution to solve such problems. Stakelberg game \cite{r140,r190,r191},  coalitional game \cite{r139,r192} and matching games \cite{r173,r193} are of the most widely used game theniques in the edge comuting papers which havs a good matching with common edge problems like resource allocation and sharing and pricing.

\subsubsection{Meta-heuristic techniques}
Metaheuristic techniques are well-known group of optimization techniques which are widely used in scientific area to fina a good enough solutions for np-hard problems. These techniques are commonly inspired by the nature and many problems can be well modeled by them. The fast convergence feature of metaheuristic algorithms makes them a good solution for time constrained edge problem such as resource management and task allocation. Metaheuristic techniques have been used in 72 articles. 30\% of the articles are fallen in the offloading topic, 51\% in the resource management topic, 18\% in architecture topic, 17\% in network topic, 20\% in security topic, 20\% in data management topic and 1\% in the economic topic. This technique, like heuristic techniques, has a high percentage of usage in resource management and offloading topics for the same reasons. Some examples of the widely used metaheuristic algorithms in edge computing area include genetic \cite{r194,r195}, ant colony \cite{r196,r197}, simulated annealing \cite{r198,r199} and particle swarm optimization \cite{r200,r201}. 

\subsubsection{Markov decision process}
The Markov Decision Process (MDP) is a mathematical technique for modeling specific types of sequential decision-making problems under uncertainty \cite{r202}. It is based on the interactions of an agent with the environment and consist of sequence of several decision problems. The Markov decision process is formally defined by 5 tuple $(S, A, T, p(), r() )$ in which S is the state space, A is the set of all possible actions, $T$ is the set of time steps in which decition-making should be made, $p()$ is the state transition probability function and $r()$ is the reward function defined on state transition. The Markov decision processes can be used to model the state evolution dynamics of a stochastic system controlling by an agent that chooses the actions at each time step.The procedure of choosing such actions is called an action policy. A policy can be history-dependent or can only consider the current state \cite{r203}. 

The advantage of MDP is that it cloud be used in both distributed and centralized scenarios. Therefore, it is suitable for modeling problems in edge computing. However, it has two problems. On the on hand, in order to reach the optimal state, decsion maker has to needs the knowledge of system state to make decisions, so it faces to the data collection problem. On the other hand, it faces the problem of curse dimensionality, meaning the state space of the corresponding dynamic problem is intractably large. So, it is no scalable and classic MDP algorithms, such as value iteration policy and policy iteration policy, are not suitable for addressing the MDP problem. In reviewed studies, methods such as RL have been used to solve this problem.

Out of 54 articles related to this technique, 48\% of the articles are fallen in the offloading topic, 42\% in the resource management topic, 18\% in architecture topic, 29\% in network topic, 7\% in security topic, 24\% in data management topic and 1\% in the economic topic. The percentage of this technique in the offloading topic is significantly more than other topics as it is a good option for modeling wide range of edge computing decision making problems raised in offloading topcs. Some example of such problems are task offloading decision-making \cite{r204}, next hob selection in a software-defined vehicular networks \cite{r205} and random access process in M2M communications \cite{r206}.

\subsubsection{Lyapunov optimization technique}
Lyapunov optimization technique is a  usefull method for optimizing some performance objectives subject to a real or virtual queue stability. This technique is a good option for solving problems dealing  with the tradeoff between the utility maximization and delay or any time-average constraints. Assuming that $Q(t) = ( Q_1(t), Q_2(t), … , Q_k(t))$ be a random vector of queues evolving over time slots. Lyapunov function is defined as $L(Q(t))$ which is a nonnegative function that tends to be larger when the system approaches to undesirable state. The changes in this function between two consecutive time slots $\Delta(t) = L(Q(t+1)) – L(Q(t))$ is defined lyaponuv drift. The queue stability can be achived by defining a convenient Lyapunov function that greedly minimizes a bound on $\Delta(t)$ every time slot.  Equation\ref{eq:1}  shows a commonly used Lyapunov function. In this equation where $b$ and w are given positive constants, with $\beta > 1$ \cite{r207}.

\begin{equation}
    \label{eq:1}
    L(Q) = \frac{1}{b} \sum_{k = 1}^{K} {W(k) {\mid Q_k \mid}^b}
\end{equation}

Edge computing problems such as resource management,  offloading and etc cloud be highly complex. Using this optimization method, the computation complexity of problems in edge computing can be greatly reduced. For this reason, the use of Lyapunov optimization in modeling edge computing problems is widely used. Although the Lyapunov method is a centralized modeling method, due to its low computation complexity, it can be used in large-scale scenarios. However, Lyapunov optimization only achieves local optimization because it is not a repetitive method and operates in a single-step optimization.

This technique is widely used in edge compuating optimization problems with different criteria. Examples of such works include minimizing power consuption subject to user quality of service constraints \cite{r208,r209,r210}, minimizing the communication resorce consumption subject to delay constraint \cite{r64}, and minimize the energy consumptions for task executions subjecting to incentive constrant to prevent the user free-riding behavior \cite{r211}.

Among all papers, 43 articles have used Lyapunov optimization technique, 54\% of the articles are fallen in the offloading topic, 64\% in the resource management topic, 14\% in architecture topic, 28\% in network topic, 4\% in security topic, 4\% in data management topic and 2\% in the economic topic. As it is clear, this technique has been used more in the resource management, offloading and network management topics, in which stabilization and load balancing of the system along with other criteria are common issues, and using this technique is a good option for optimizing such problems.

\subsubsection{Graph theory}
The graph theory is related to studying mathematical problems dealing with graphs and has applications in many scientific and engineering issues. The graph-based thechniques can be a good option to solve any problem that can be modeled by some objects and relationship between them \cite{r212}. These techniques are also of the most widely used techniques in edge computing and in many articles the edge computing platform is modeled as a graph in which nodes represent computing server and edges represent the links between them \cite{r213,r214}. Also some applications in this field can be represented by graph in which node represents functions or tasks and edges represent dependency between these tasks \cite{r33,r215}. Among the 41 articles related to graph techniques, 20\% of the articles are fallen in the offloading topic, 60\% in the resource management topic, 20\% in architecture topic, 20\% in network topic, 15\% in security topic, 20\% in data management topic and 7\% in the economic topic. As it is clear from the statistics in Figure 9, the use of graph theory techniques in resource management is significantly more than other topics. This is because resources in edge computing are distributed and are often modeled as graphs. The mostly used graph theory techniques int the reviewd papres are graph coloring \cite{r159,r216} graph matching \cite{r217,r218} and partitioning \cite{r215,r219}.

\subsubsection{Learning technique}
Learning techniques are about giving computers the ability to learn from the data by themselves without being explicitly programmed \cite{r220}. These techniques focus on automatically improving computer systems through experience \cite{r221}. Machine learning techniques are suitable for dealing with high complexity problems. Therefore, machine learning methods, i.e. as reinforcement learning, can overcome the issue of the curse of dimensionality in MDP.  Also, These approaches (especially online learning ones) are very suitable for dynamic environments and can also be applied to centralized and or distributed applications. Edge computing environment is completely dynamic due to user mobility and workload changes. Therefore, the use of learning methods to predict the state of the environment and make an intelligent decision has been considered by researchers. Also, one of the important applications of learning methods is extracting information from raw data. This problem is very essential in edge computing, because the applications developed at the edge and IoT devices are constantly generating raw data. This data has little value until it is analyzed. Therefore, this case is one of the reasons for the widespread use of learning techniques in papers. 

This technique has been used in 210 articles, among which 24\% fallen in the offloading topic, 40\% in the resource management topic, 25\% in architecture topic, 24\% in network topic, 16\% in security topic, 37\% in data management topic and 1\% in the economic topic. One of the notable points in this statistic is the greater usage of this technique in data management articles. This is because these techniques are closely related to data both to discover new information and to improve learning and accuracy. Reinforcement learning \cite{r222,r223}, neural network \cite{r224,r225} and clustering \cite{r110,r226} are of the most widely used learning techniques in the edge computing literature. 

\subsection{Which forms of empirical evaluation have been used?}
\begin{figure}[t]
	\centering
	\includegraphics[scale=0.51]{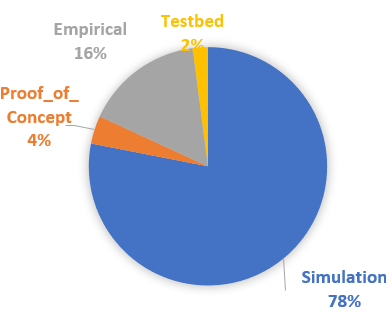}
	\caption{Evaluation Methods in Edge Computing}
	\label{fig:11}
\end{figure}

Figure\ref{fig:11} shows the statistics of different evaluation methods in the reviewed articles. Common methods seen in papers for evaluation include simulation methods, empirical, implementation on a testbed, and proof-concept methods. As it is clear, most of the proposed methods have been tested and evaluated by simulation due to the unavailability of a real edge infrastructure or high cost and complexity of real world evaluations. FogNetSim \cite{r227}, iFogSim \cite{r228}, EdgeCloudSim \cite{r229}, IoTsim \cite{r230} are examples of simulation tools in edge and cloud computing. The empirical evaluation consists of different kind of approaches, such as developing prototypes (21\%) and scenarios (4\%), designing a case study (12\%), utilizing a real dataset (5\%), using benchmarks (5\%), use cases (5\%) and platforms (1\%).

\subsection{Which qualitative requirements have been considered to move towards edge computing?}

\begin{figure}[t]
	\centering
	\includegraphics[scale=0.51]{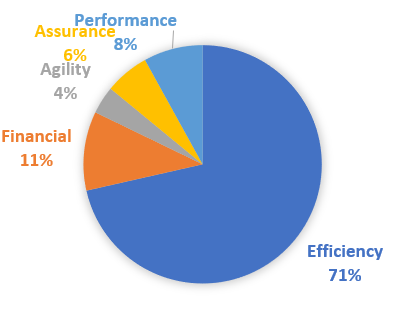}
	\caption{Frequent QoS Parameters in Edge Computing}
	\label{fig:12}
\end{figure}

As shown in Figure\ref{fig:12}, different QoS categories are frequently reported among the reviewed papers. This section provides statistics on the main QoS category based on SMI \cite{r231} and ISO 9126 \cite{r232} quality models related to edge environment generally. 
The efficiency category has the most repetition among the studies. Efficiency is defined as "A set of attributes that bear on the relationship between the level of performance of the software and the amount of resources used, under stated conditions \cite{r232}". Time (49\%), energy consumption (26\%), utilization (22\%), and throughput (3\%) had the most frequency among the efficiency criteria. 

As mentioned earlier, applications in edge computing environment are latency-sensitive applications. Reducing this delay has been the starting point of their migration from the cloud to the edge of network. Therefore, it is natural that the time criterion specially delay and latency should be considered as the main one in evaluating the QoS of such programs. 

After the efficiency category, the financial category has the second rank. Financial category focuses on "The amount of money spent on the service by the client" \cite{r231}. Cost is considered as the only criterion in this category. Performance category focuses on "attributes of the features and functions of the provided service" \cite{r231}. 52\% of studies only reported performance as their quality criterion without any mention to its sub-criteria. Some studies reported accuracy (48\%) as a performance criterion in their evaluation. 

Criteria related to assurance category is the next. Assurance is considered as "attributes that indicate how likely it is that the service will be available as specified" \cite{r231}. Reliability (50\%), Availability (26\%), Failure (1\%), Stability (19\%) and Robustness (4\%) are the most frequent criterion in assurance category.

Agility is defined as "attributes including the impact of a service upon a client 's ability to change direction, strategy, or tactics quickly and with minimal disruption" \cite{r231}. Agility parameters such as flexibility (29\%), elasticity (2\%) is reported frequently in studies.

\section{Comparision with Similar Review Researches}
As mentioned before, at the beginning of the search process, several secondary studies (survey, reviews) were selected randomly as the starting point by the expert. Furthermore, during the search process, other review studies were found. A list of these studies is provided in [SuppFil]\_(E2,T2). Among the review studies, studies specific to a particular sub-topic were excluded. For the remaining studies, covered topics in the field of edge computing were extracted that details of this information are provided in [SuppFile]\_(E2,T6) . Table\ref{tbl6} shows the comparison between our SMS and other review studies. In contrast to our SMS, none of these studies used a systematic approach to cover all related journal studies in the field of edge computing. However, we selected six papers \cite{r233}, \cite{r2}, \cite{r1}, \cite{r234}, \cite{r235}, \cite{r40}, and \cite{r236} for comparison, which are relatively more comprehensive and address more aspects of the field of edge computing.

The differences between these studies and our SMS are presented in Table\ref{tbl7}. For each of the studies being compared, the necessary information about the number of covered venues and studies have been extracted. Table 8 shows that our SMS (last row) reviewed and covered an extensive collection of related studies compared to other reviews. For example, in review paper \cite{r2}, out of 450 studies cited, only 120 papers are used in taxonomy. Also, these studies in the field of edge computing have not undergone a systematic process and address fewer topics and sub-topics (see [SuppFile]\_(E2,T3)). Unlike these reviews, our SMS uses a comprehensive and complete process to find related studies in a collection of high-quality journals.

\begin{table*}[h]
		\caption{Comparison Between Our SMS and Other Related Reviews.}\label{tbl6}
		\begin{tabular*}{\textwidth}{@{\extracolsep{\fill}} p{2cm} lccccccc 
		@{\extracolsep{\fill}} }
			\toprule
			\diagbox{PID}{Topic} & Scope  & Ar. & NM & DM  & RM & Ec. & CO  & S\&P  \\
			\midrule
		     68        &  FOG  &    \cmark   &  \xmark & \xmark & \cmark & \xmark & \cmark & \xmark   \\
		     1610      & MEC   &    \cmark   &  \xmark & \xmark & \cmark & \xmark & \cmark & \xmark \\
		     1611      & EC    &    \cmark   &  \xmark & \xmark & \xmark & \xmark & \xmark & \cmark \\
		     1618      & MEC   &    \cmark   &  \xmark & \xmark & \cmark & \xmark & \cmark & \xmark \\
		     5439      & FOG   &    \cmark   &  \cmark & \xmark & \cmark & \xmark & \xmark & \xmark \\
		     6838      & EC    &    \cmark   &  \xmark & \xmark & \xmark & \xmark & \xmark & \xmark \\
		     5715      & EC    &    \cmark   &  \xmark & \cmark & \cmark & \xmark & \cmark & \xmark \\
		     7906      & FOG   &    \cmark   &  \xmark & \xmark & \cmark & \xmark & \cmark & \cmark \\
		     7907      & FOG   &    \xmark   &  \cmark & \xmark & \cmark & \xmark & \xmark & \cmark \\
		     8044      & EC    &     \xmark  &  \cmark & \xmark & \cmark & \xmark & \xmark & \xmark \\
		     8242      & EC    &    \cmark   &  \xmark & \cmark & \xmark & \xmark & \cmark & \cmark \\
		     8587      & EC    &    \cmark   &  \xmark & \xmark & \cmark & \xmark & \cmark & \xmark \\
		     8775      & EC    &    \cmark   &  \xmark & \xmark & \cmark & \xmark & \xmark&  \cmark \\
		     This SMS  & EC    &    \cmark   &  \cmark & \cmark & \cmark & \cmark & \cmark & \cmark \\ 
			\bottomrule
		\end{tabular*}
	\end{table*}
	
	\begin{table*}[h]
	\begin{center}
	\begin{minipage}{\textwidth}
	\caption{Statistical Comparison between Our SMS and Other Selected Reviews.}\label{tbl7}
		\begin{tabular*}{\textwidth}{@{\extracolsep{\fill}} 
		p{0.5cm} p{0.8cm} p{1.1cm} p{0.5cm} p{1.5cm} p{0.3cm} p{0.3cm} p{2cm} 
		@{\extracolsep{\fill}} }
			
			\toprule
			Ref.    & Type   &	Studies\# & Venue\#	& Searching Method & QC\footnotemark[1] &	Eval.\footnotemark[2] &	Period \\
			\midrule
			{\cite{r225}}   & Survey &	22/169	  &  41	& NM	                         & \xmark &	\xmark &	2013-2017  \\
			{\cite{r52} }   & Survey &	120/450	  &  40	& Manual 	                     & \xmark &	\xmark &	2013-2017 \\
			{\cite{r1}  }   & Survey &	37/137	  &  29	& NM	                         & \xmark &	\xmark &	2004-2017 \\
			{\cite{r226}}   & Survey &	27/99	  &  14	& NM	                         & \xmark &	\xmark &	2002-2019 \\
			{\cite{r227}}   & Survey &	97/296	  &  55	& NM	                         & \xmark &	\xmark &	2002-2019 \\
			{\cite{r42} }   & Survey &	70/200	  &  43	& NM	                         & \xmark &	\xmark &	2009-2019 \\
			{\cite{r228}}   & Survey &	36/101	  &  19	& NM	                         & \xmark &	\xmark &	2009-2019 \\
			Our SMS	& SMS	 &  1440/1440 &	112	& Manual-Snowballing-Database	& \cmark &	\cmark &	2001-2019  \\ 
		
		\end{tabular*}
		\footnotetext[1]{QC(Quality Criteria)} 
        \footnotetext[2]{Eval. (Search Evaluation)}
		\end{minipage}
		\end{center}
        
	\end{table*}

\section{Implications of Findings}
In this paper, we have conducted a systematic mapping study (SMS) to create a guide for researchers, practitioners, and professors working in the field of edge computing. The findings of this paper can be useful for different groups of audiences. It is expected that most of the readers of this paper will be from three general categories such as researchers, practitioners working in the industry, and university professors active in this field. Therefore, in this section, we have provided tips and guidelines for each of these three main groups of audiences.

\subsection{Implications for researchers}
The field of edge computing has become one of the most popular and prolific fields in recent years, and extensive researches has been done on it.  In this SMS, active research topics in the field of edge computing have been identified by reviewing a large number of articles in this field and it can be a suitable starting point for new researchers interested in edge computing (RQ3). There is also a significant difference in the number of studies conducted between different countries (RQ2). China (31\%), the United States (11\%), and Canada (7\%) account for the highest percentage of published articles among countries. In these countries, technological advances in the industry compared to other countries and the deep connection between industry and academia. The pioneer reasercher of this field in each topic are also identified in this SMS (RQ1). This information can be useful for reaserchers who want to have international collaborations.

Over the past decade, the field of edge computing has made significant progress, and various architectures for the use of edge resources have been proposed (RQ5).  These architectures may be misunderstood by some reaserchers or used interchangeably in the litrature. One of the necessary things for researchers in this field is complete familiarity with standard edge architectures and paying attention to the differences and characteristics of each of them. This knowledge helps researchers to understand the basic concepts deeply and proposeeffective and accurate solutions. 

The evaluation methods used in most of the reviewed papers in the field of edge computing are based on simulation (RQ7). Since the foundations for the full realization of edge computing infrastructure are not yet available in many parts of the world; the use of simulation can be a simple and inexpensive alternative to evaluate researches in this field. Due to this issue, so far, various simulators for edge computation work have been proposed, such as FogNetSim++ \cite{r227} and iFogSim \cite{r228} . Although simulation is a low-cost and hassle-free way to evaluate a research, but simulations are usually accompanied by simplifications and hypotheses that may be far from the truth. Therefore, the implementation of real testbeds and small prototypes at the laboratory level can help to more accurately assess and identify operational challenges and problems in this area. For example, the Raspberry Pi can be used to build an almost realistic platform to create the scenarios desired by the researchers \cite{r237}. Therefore, the researchers can move towards the use of real implementation for more accurate assessments and identification of operational challenges.

Another goal in this SMS was identifing and introducing common and widely used applications and techniques in the edge computing papers. Optimization methods are widely used in the literatures of this field for load balancing and increasing the efficiency. Therefore, familiarity with the methods and techniques of machine learning, optimization, meta-heuristics, or game theory can be very handy and practical for researchers in this field. Frequent applications in edge computing papers were also explored in a separate section Researchers can use the information of these sections to get acquainted with the common techniques and applications in this field  (RQ6). Familiarity with these applications and techniques as a prerequisite makes it easier for researchers to read and undrstand the articles of this field.

\subsection{Implications for practitioners}
After years of research, the field of edge computing is now approaching its maturity. In the academic field, a lot of researches has been done to exploit resources close to users, and a variety of ideas have been presented to overcome the various challenges in this field (RQ2). Therefore, a wide scientific source of solutions, ideas and applications in the academic field has been provided. Utilizing this source of knowledge to solve the real-world problems and increase the quality of daily life, requires the connection of industrial organizations with scientific and academic departments. This connection can be made through participation in conferences related to the field of edge computing in order to transfer knowledge from researchers to industrial people, as well as transfer challenges and operational problems from people working in the industry to researchers. This connection can accelerate the progress of realization of edge computing schemes in the real world.

With the advent of IoT, new applications have been introduced, some of which cannot be implemented without the use of edge resources (RQ4). Since these applications are expected to become part of people's daily lives in the future world, the creation and development of edge computing platforms will be one of the necessities of the future world. Therefore, developers of applications and services can pay attention to the use of edge resources and distributed platforms near the user in their designs. Also, data network operators and network equipment manufacturers need to consider the realization of the edge computing model in the design of future network infrastructure or communication equipment. 

\subsection{Implications for teachers}
In the last decade, edge computing has been considered as a promising solution to increase service quality and reduce response time by bringing resources closer to users' devices. Also, in recent years, there has been significant growth in researches in this field so that this field of research can be considered as a broad, independent, and popular field among researchers (RQ2). Therefore, it is vital to acquaint students and novices (in academia or industry) with the concepts of edge computing to transfer knowledge and research achievements of recent years. Professors active in the field of cloud computing and computer networks can include the concepts in this field as the main topics in courses such as "cloud computing" and "distributed systems" or implicitly refer to it in related courses such as "computer networks" and "internet engineering". Professors can also use the concepts presented in the section (RQ3) and statistical results obtained from the papers reviewed in the sections (RQ1, RQ3) to get acquainted with the concepts related to this field and prepare educational content.

\section{CONCLUSION}
With the increasing trend towards edge computing and the dramatic increase in the number of studies in this field, the need for a systematic review in this field was felt. In this paper, we conducted an SMS in the field of edge computing from 2001 to 2020. For this aim, we used well-known methodologies for systematic review as well as our past experiences in publishing systematic reviews. The main part of the SMS methodology used involves providing a robust search method. For this purpose, the search was performed at three levels (manual search, backward snowballing search, and database search). We defined criteria for selecting studies with the required quality and maximum relevance to the field of edge computing. Based on these criteria out of 8725 obtained related study, 1440 studies were selected. We also evaluated the search methodology to ensure adequate coverage of the studies.

According to the methodology used, we designed 8 research questions and answered the questions during the SMS. The first three questions include introducing the most important researchers and pioneers (RQ1), reviewing the number of researches done (RQ2), and extracting the most important topics and sub-topics (RQ3) in the field of edge computing. In the next three questions (RQ4, RQ5, and RQ6), the number of researches in the field of edge computing has been examined and analyzed from the perspective of extractive topics. In these three questions, the most crucial applications of edge computing in each topic, the most important architectures used in each topic, and the techniques used in each topic are reviewed and analyzed, respectively. The most important evaluation methods and quality criteria used in the research to move towards edge computing are reviewed in questions RQ7 and RQ8, respectively. Besides, we compared our SMS with several reviews in the field of edge computing and examined the depth of the method used and the various aspects considered.

Our research shows that edge computing is an active and growing research field in different geographical areas. Although the number of researches in this field has increased significantly, there are still challenges in this field, and the development of methods and paradigms is needed to solve these challenges. Systematic review studies such as our SMS can be used to provide more detailed review of each aspect using the review systematic literature (SLR). For future work, each of the high-level topics extracted in this paper can be explored in-depth to answer more specific research questions.

\bibliography{edge-bibliography}


\section*{Statements and Declarations}
\subsection*{Funding} The authors acknowledge the funding of Cloud Computing Lab (CCLab), Department of Computer Engineering, Ferdowsi University of Mashhad, Iran. 
\subsection*{Author Contributions}
All authors contributed to the study conception and design. Material preparation, data collection were performed by [Sakhdari, Zolfaghari, Izadpanah, Mehdizadeh, Shadi, and Rahati], and analysis were performed by [Sakhdari, Zolfaghari, Izadpanah, Abrishami, and Rasoolzadeghan]. The first draft of the manuscript was written by [Sakhdari, Zolfaghari, and Izadpanah] and all authors commented on previous versions of the manuscript. [Sakhdari, Zolfaghari Izadpanah,  Abrishami, and Rasoolzadegan] read and approved the final manuscript.
\subsection*{Competing Interests} The authors declare no conflict of interest. 
 \subsection*{Data availability} The datasets generated during and/or analysed during the current study are available in the github repository, \url{https://github.com/jalalsakhdari/SMS} 
\end{document}